\documentclass[aps,showpacs,preprintnumbers,amsmath, amssymb]{revtex4-2}

\usepackage{amsmath}
\usepackage{float}
\usepackage{graphics,epsfig,hyperref}
\usepackage{graphicx}
\usepackage{dcolumn}
\usepackage{bm}
\usepackage{epstopdf}
\usepackage{subfigure}

\def \be {\begin{equation}}
	\def \ee {\end{equation}}
\def \bea {\begin{eqnarray}}
	\def \eea {\end{eqnarray}}

\begin{document}
\baselineskip=0.8 cm
\title{\bf Chaotic motion of particles around a dyonic Kerr-Newman black hole immersed in the Melvin-swirling universe}
\author{Deshui Cao$^{1}$, Lina Zhang$^{1}$, Songbai Chen$^{1,2}$\footnote{csb3752@hunnu.edu.cn}, Qiyuan Pan$^{1,2}$\footnote{panqiyuan@hunnu.edu.cn}, and Jiliang Jing$^{1,2}$ \footnote{jljing@hunnu.edu.cn}}	
	
\affiliation{$^1$Department of Physics, Institute of Interdisciplinary Studies, Key Laboratory of Low Dimensional Quantum Structures and Quantum Control of Ministry of Education, Synergetic Innovation Center for Quantum Effects and Applications, and Hunan Research Center of the Basic Discipline for Quantum Effects and Quantum Technologies, Hunan Normal University,  Changsha, Hunan 410081, People's Republic of China
\\
$^2$Center for Gravitation and Cosmology, College of Physical Science and Technology, Yangzhou University, Yangzhou 225009, People's Republic of China}

\begin{abstract}
\baselineskip=0.6 cm
\begin{center}
{\bf Abstract}
\end{center}

We employ the Poincar\'{e} section, fast Lyapunov indicator, recurrence analysis, bifurcation diagram and basins of attraction to investigate the dynamical behaviors of the motion of particles around a new dyonic Kerr-Newman black hole immersed in the Melvin-swirling universe presented in [A. Di Pinto, S. Klemm, and A. Vigan\`o, J. High Energy Phys. {\bf 06}, 150 (2025)]. We note that the swirling parameter $j$ and magnetic field strength $B$ make the equations of motion for particles nonseparable, and confirm the presence of chaotic behavior in the motion in this dyonic Kerr-Newman-Melvin-swirling spacetime and its sub-cases by removing the conical singularities and removing both the conical singularities and the Dirac strings. We observe that both the number of chaotic orbits and the chaotic region increase with the increase of the parameters $j$ and $B$, but decrease as the electric charge $Q$, magnetic charge $H$ or spin parameter $a$ increases. Moreover, we find that the presence of $j$ changes the ranges of $B$, $Q$, $H$ and $a$ where the chaotic motion appears for particles. The swirling parameter together with the magnetic field strength, electric charge, magnetic charge and spin parameter yields richer physics in the motion of particles for the spacetime of a dyonic Kerr-Newman black hole immersed in the Melvin-swirling universe.

\end{abstract}

\maketitle
\newpage

\section{Introduction}

As a fascinating phenomenon in dynamical systems, the chaos is a kind of unpredictable and seemingly random motion highly sensitive to initial conditions but often exhibits underlying features such as the self-similarity \cite{Brown1996,Brown1998}.
Such behavior typically emerges in nonlinear systems, where tiny differences in initial conditions can grow rapidly at exponential rates in the chaotic motion \cite{Ott2002,Sprott2003}. Considering that the chaotic motion is very common in nature, there has been accumulated interest in studying the chaos in physical systems \cite{VarvoglisP,Letelier-Vieira,Hashimoto,DaluiMM,Falco-Borrelli1,Falco-Borrelli2,Zhang2022}.

In black hole physics, the sustained research interest has been devoted to studying the chaos. For the general Kerr-Newman spacetime, the geodesic motion of a neutral test particle is integrable and regular  \cite{Carter1968}. However, the chaos emerges in systems where the spacetime possesses complex geometrical structures or when extra interactions are introduced. Considering the trajectories of test particles for the Ernst spacetime in Ref. \cite{Karas} and later generalizing to the neutral and charged particles for a magnetized Ernst–Schwarzschild spacetime in Ref. \cite{Ldan}, the authors found that the motion of particles appears to be chaotic. Then, the chaotic phenomena have been found to exist in the perturbed Schwarzschild spacetime  \cite{Bombellitf-Calzetta,SotaSM,WitzanySS},  multi-black hole spacetimes \cite{DettmannFC}, the non-standard Kerr black hole spacetime described by Manko-Novikov metric \cite{non-standardKerr1,non-standardKerr2,non-standardKerr3,non-standardKerr4,non-standardKerr5}, the accelerating and rotating black hole spacetime \cite{schen1}, and the disformal rotating black-hole spacetime \cite{ZhouSCPMA}. Specifically, we investigated the motion of particles in the
spacetime of a Kerr black hole immersed in swirling universes and confirmed the presence of chaos in the motion of particles
in this background spacetime \cite{Cao}. Extending the investigation to extra interactions, such as the Einstein tensor coupling in Schwarzschild-Melvin spacetime \cite{mschen2}, the Chern-Simons invariant coupling in Kerr spacetime \cite{Zhou} and Einstein-Maxwell dilaton spacetime \cite{Zhang}, the researchers also observed the chaotic dynamics of test particles. Additionally, there exist the chaotic behaviors in string dynamics in the Schwarzschild black hole spacetime \cite{FrolovL} and in AdS black hole spacetimes \cite{ZayasT,Basu,MaWZ}, and the thermal chaos in the extended phase space \cite{ChaosExtendedPhaseSpace1,ChaosExtendedPhaseSpace2,ChaosExtendedPhaseSpace3,ChaosExtendedPhaseSpace4}.

It is of great interest to generalize the investigation on the dynamical behaviors of the motion of particles around a new dyonic Kerr-Newman black hole immersed in the Melvin-swirling universe presented by Di Pinto \emph{et al.} in Ref. \cite{DiPinto} more recently, which is an exact solution of Einstein-Maxwell field equations representing a rotating black hole with both electric and magnetic charges immersed in a universe that itself is also rotating and magnetized. Compared to the Kerr black hole immersed in swirling universes \cite{Astorino2} (for the
related studies see \cite{Barrientostqb,BarrientosCKMOP,Capobianco,MoreiraHC,ChenCJ,BarrientosCHP,Gjorgjieski,Gu,Ouyang}), this dyonic Kerr-Newman black hole is a quite general solution characterized by six parameters: the mass $M$, the charges $Q$ and $H$, the angular momentum $a$, the external magnetic field $B$ and the swirling parameter $j$, which encompasses many interesting sub-cases. Thus, the motivation for completing this work is two-fold. On one level, it is worthwhile to study the effects of the combination of the swirling parameter $j$ and the magnetic field strength $B$, electric charge $Q$, magnetic charge $H$, spin parameter $a$ on the motion of particles and ask whether there exists chaos in this dyonic Kerr-Newman black hole spacetime. On another more speculative level, it would be important to systematically compare with the findings shown in Ref. \cite{Cao} for the spacetime of a Kerr black hole immersed in swirling universes and in Ref. \cite{Yang,Liu} for magnetized Kerr–Newman spacetimes, and see some general features for the chaotic motion of particles in the spacetime of a dyonic Kerr-Newman black hole immersed in the Melvin-swirling universe.

The organization of this work is as follows. In Section II, we introduce the new dyonic Kerr-Newman black hole immersed in a Melvin-swirling universe and present the corresponding geodesic equations of particles. In Section III, we use the techniques including the Poincar\'{e} section, fast Lyapunov indicator  (FLI), recurrence analysis, bifurcation diagram and basins of attraction to investigate the chaotic motion of particles around this dyonic Kerr-Newman black hole immersed in the Melvin-swirling universe. We summarize our results in the last section. We extend the study, in the Appendix, to the chaotic motion of particles in sub-cases of the dyonic Kerr-Newman-Melvin-swirling spacetime by removing the conical singularities and removing both the conical singularities and the Dirac strings.
	
\section{Geodesics of particles}
	
The dyonic Kerr-Newman black hole in a Melvin-swirling universe is the exact solution of Einstein-Maxwell field equations, which represents a rotating black hole with both electric and magnetic charges immersed in a universe that itself is also rotating and magnetized. The corresponding line element expressed in Boyer-Lindquist coordinates is given by \cite{DiPinto}
\begin{equation}\label{metric}
	\begin{split}
	ds^{2} &=F \biggl(-\frac{\Delta}{\Sigma} dt^{2}+\frac{dr^{2}}{\Delta}+d\theta^{2}\biggr)+\frac{\Sigma \sin^{2}\theta}{F} \biggl(d\varphi-\frac{\Omega}{\Sigma} dt\biggr)^{2}, \\
	A &= \frac{A_0}{\Sigma} \,dt + \frac{A_{3}}{F}\biggl( d\varphi - \frac{\Omega}{\Sigma}\,dt \biggr),
\end{split}
\end{equation}
where the metric functions are defined by
\begin{align}
	F &= R^{2}+2 B \phi_{(0)}+\frac{B^{2}}{2} \phi_{(1)}+\frac{B^{3}}{2} \phi_{(2)}+\biggl(j^{2}+\frac{B^{4}}{16} \biggr) \phi_{(3)}+2 j B \phi_{(4)}+j F_{(1)}\,, \\
	\Omega &= a \lambda+2 B \chi_{(0)}+\frac{B^{2}}{2} \chi_{(1)}+\frac{B^{3}}{2} \chi_{(2)}+\biggl(j^{2}+\frac{B^{4}}{16} \biggr) \chi_{(3)}+2 j B \chi_{(4)}+j \Omega_{(1)}\,, \\
	A_0 &= \chi_{(0)} + \frac{B}{2}\,\chi_{(1)} + \frac{3B^2}{4}\,\chi_{(2)} + \frac{B^3}{8}\,\chi_{(3)} + j \chi_{(4)} \,, \\
	A_3 &= \phi_{(0)} +  \frac{B}{2}\,\phi_{(1)} + \frac{3B^2}{4}\,\phi_{(2)} + \frac{B^3}{8}\,\phi_{(3)} + j \phi_{(4)} \,,	
\end{align}
and 
\begin{align}
	\Delta & = r^2 - 2Mr + Z^2 + a^2 \,, \qquad\qquad\qquad  Z^2 =Q^2+H^2\,, \\   
	\Xi &= \bigl(r^2 + a^2\bigr )\sin^2\theta + Z^2\cos^2\theta \,,\qquad\quad  R^2 = r^2 + a^2\cos^2\theta \,,
	 \\
	\label{def-sigma}
	\Sigma & = \bigl(r^2 + a^2\bigr)^2 - \Delta a^2 \sin^2\theta \,, \qquad\qquad\;\;
	\lambda = r^2 + a^2 - \Delta = 2Mr - Z^2 \,,
\end{align}
with the expansion coefficients
\begin{subequations}
	\begin{align}
		\phi_{(0)} & = a Q r \sin^2\theta - H \bigl( r^2 + a^2 \bigr) \cos\theta \,, \\
		\phi_{(1)} & = \Sigma \sin^2 \theta + 3 Z^2 \bigl( r ^2\cos^2\theta + a^2 \bigr) \,, \\
		\begin{split}
			\phi_{(2)} & = a Q \biggl[\bigl( 1 + \cos^2\theta \bigr) \Bigl( r^3 +  \bigl (2 M + r \bigr) a^2 \Bigr) + r \cos^2\theta \Bigl(2 Z^2 - \Delta \bigl(3 - \cos^2\theta \bigr) \Bigr)\biggr] \\
			&\quad + H \cos\theta \Bigl[ 2 a^2 \lambda \sin^2\theta - \bigl(r^2 +a^2) \, \Xi \Bigr] \,,
		\end{split}
		\\
		\begin{split}
			\phi_{(3)} & = Z^2 \biggl[2 a^4 \bigl( 1 + \cos^2\theta \bigr)^2 +  r^2 \cos^2\theta \bigl( \, \Xi + R^2 \sin^2\theta \bigr) \\
			&\quad + a^2 \cos^2\theta \Bigl( 2\, \Xi + 3 Z^2 + r^2 \bigl( 5 + 6 \sin^2\theta + 3 \cos^4\theta \bigr) - 8 \Delta \Bigr) \biggr] \\
			&\quad + a^2 \Bigl( \lambda^2 \cos^2\theta \bigl( 3 - \cos^2\theta \bigr)^2  + r^3 \sin^6\theta ( 4 M - r )\Bigr) + 2 a^4 \Bigl( 2 M^2 \bigl( 1 + \cos^2\theta \bigr)^2 - \Delta \sin^6\theta \Bigr) \\
			&\quad + a^6 \sin^6\theta + \bigl( r^2 + a^2 \bigr)^3 \sin^4\theta \,,
		\end{split}
		\\
		\phi_{(4)} & = a H \biggl[2 M \Bigl(a^2 +  \cos^2\theta \bigl(2 r^2 + a^2 \bigr) \Bigr) + r \sin^2\theta \bigl( \lambda + \Delta \sin^2\theta \bigr)\biggr] + Q \cos\theta \Bigl[ \bigl( r^2+ a^2 \bigr)\, \Xi - 2 \lambda a^2 \sin^2\theta \Bigr] \,,
	\end{align}
\end{subequations}
and
\begin{subequations}
	\begin{align}
	\chi_{(0)} & = a H\Delta \cos \theta - Q r \bigl(r^2+a^2\bigr) \,, \\
	\chi_{(1)} & = -3 a Z^2 \Bigl( \lambda + \Delta \bigl(1+\cos^2\theta \bigr) \Bigr) \,, \\
	\begin{split}
	\chi_{(2)} & = Q \biggl[ r^3 \Bigl( \lambda + \Delta \bigl(1+\cos^2\theta \bigr) \Bigr) + a^2 \Bigl( \Delta \cos^2\theta \bigl( 3r - 4 M \bigr) -  r \bigl( Z^2 + \Delta \bigr)   \Bigr) - 2 M a^4 \biggr] \\
	&\quad + a H \Delta \cos\theta  \bigl( \, \Xi + 2 R^2 \bigr)  \,,
	\end{split}
	\\
	\begin{split}
	\chi_{(3)} & = a \biggl[6 M r^5 - a^2 \Delta \cos^2\theta \Bigl ( \bigl( Z^2 + 4 M^2 - 6 M r \bigr)\cos^2\theta + Z^2 + 12 M^2 - 12 M r - 6 r^2 \Bigr) - 2 a^4 M \bigl( 2 M + r \bigr) \\
	& \quad - a^2 Z^2\Delta + 4 a^2 M r \bigl( r^2 - 2 Z^2 + 3 M r \bigr) - \Delta \cos^2\theta \Bigl( 6 r^2 \bigl(\Delta - r^2 \bigr) + \bigl( Z^4 + 2 M r^3 - 3 Z^2 r^2 \bigr) \cos^2\theta \Bigr)\biggr] \,,
	\end{split}
	\\
	\begin{split}
	\chi_{(4)} & = - a Q \Delta \cos\theta \bigl( \, \Xi + 2 R^2 \bigr) \\
	&\quad + H \biggl[ r^3 \bigl( \Delta \cos^2\theta + r^2 \bigr) - a^2 \Bigl(r \bigl( Z^2 - r^2 \bigr) + 4 M \Delta \cos^2\theta + r \Delta \bigl( 1 - 3 \cos^2\theta \bigr) \Bigr) - 2 a^4 M\biggr] \,,
	\end{split}
	\end{align}
\end{subequations}
and finally
\begin{subequations}
	\begin{align}
	F_{(1)} &=-4 a \cos\theta \Bigl[M \bigl(1+\cos^{2}\theta \bigr) \bigl(r^{2}+a^{2} \bigr)+\lambda r \sin^{2}\theta \Bigr] \,, \\
	\Omega_{(1)} &=-4 \Delta \cos \theta \biggl[r^{3}+a^{2} \Bigl(\bigl(r-M\bigr) \cos^{2} \theta-M \Bigr)\biggr] \,.
	\end{align}
\end{subequations}
This black hole is characterized by its mass $M$, electric charge $Q$, magnetic charge $H$ and angular momentum per unit mass $a$, and is embedded in an external magnetic Melvin field of strength $B$, with the swirling parameter $j$ corresponding to the rotation of the background spacetime. The metric reduces to a Kerr black hole embedded in swirling universes described in \cite{Astorino2} when $B = Q = H = 0$ and to the dyonic Kerr-Newman-Melvin metric \cite{Gibbons} when the swirling parameter $j=0$. The curvature singularity occurs at $F = 0$, while the coordinate singularities remain unaffected by the swirling and Melvin parameters. The spacetime has two event horizons, located at
\begin{equation}
r_{\pm} = M\pm\sqrt{M^2-a^2-Z^2},
\end{equation}
corresponding to $\Delta = 0$. Di Pinto \emph{et al.} \cite{DiPinto} found that, with the suitable choice of the parameters, the metric can be free not only of string singularities but also of curvature singularities, which leads to a spacetime smooth everywhere. The geometric units $(G = c = 1)$ are adopted throughout this work.

In order to obtain the geodesics in the spacetime (\ref{metric}), the Lagrangian of a particle is given by
\begin{equation}
\mathcal{L}=\frac{1}{2}g_{\mu\nu}\dot{x}^{\mu}\dot{x}^{\nu}=\frac{1}{2}\left(g_{t t} \dot{t}^{2}+g_{r r} \dot{r}^{2}+g_{\theta \theta} \dot{\theta}^{2}+g_{\varphi \varphi} \dot{\varphi}^{2}+2 g_{t \varphi} \dot{t} \dot{\varphi}\right),
\end{equation}
where $\dot{x}^{\mu} = \frac{dx^{\mu}}{d\tau}$ with a proper time $\tau$. From the definition of the generalized momentum $p_{\mu} = \frac{\partial \mathcal{L}}{\partial \dot{x}^{\mu}}$, we have
\begin{align}
p_{t}= g_{tt}\dot{t}+g_{t\varphi}\dot{\varphi} \equiv -E, \quad
p_{\varphi}= g_{\varphi\varphi}\dot{\varphi}+g_{t\varphi}\dot{t} \equiv L,
\end{align}
where $E$ and $L$ denote the energy and angular momentum of the particle, respectively. The equations of motion for the particle are
\begin{equation}\label{wfE1}
	\dot{t}=\frac{{g}_{\varphi\varphi} E+{g}_{t \varphi} L}{{g}_{t \varphi}^{2}-{g}_{t t} {g}_{\varphi\varphi}},\quad
	\dot{\varphi}=-\frac{{g}_{t \varphi} E+{g}_{t t} L}{{g}_{t \varphi}^{2}-{g}_{t t} {g}_{\varphi\varphi}},
\end{equation}
and
\begin{equation}\label{wfE2}
	\ddot{r}=\frac{1}{2} g^{rr}\left(g_{t t, r} \dot{t}^{2}-g_{r r, r} \dot{r}^{2}+g_{\theta \theta, r} \dot{\theta}^{2}+g_{\varphi\varphi, r} \dot{\varphi}^{2}+2 g_{t \varphi, r} \dot{t} \dot{\varphi}-2 g_{rr, \theta} \dot{r} \dot{\theta}\right),
\end{equation}
\begin{equation}\label{wfE3}
	\ddot{\theta}=\frac{1}{2} g^{\theta \theta}\left(g_{t t, \theta} \dot{t}^{2}+g_{r r, \theta} \dot{r}^{2}-g_{\theta \theta, \theta} \dot{\theta}^{2}+g_{\varphi\varphi, \theta} \dot{\varphi}^{2}+2 g_{t \varphi,\theta} \dot{t} \dot{\varphi}-2 g_{\theta \theta, r} \dot{r} \dot{\theta}\right).
\end{equation}
Additionally, the motion of the particle is constrained by the normalization condition of its four-velocity
\begin{equation}\label{Hcon}
	h=g_{t t} \dot{t}^{2}+g_{r r} \dot{r}^{2}+g_{\theta \theta} \dot{\theta}^{2}+g_{\varphi\varphi} \dot{\varphi}^{2}+2 g_{t \varphi} \dot{t} \dot{\varphi}+1=0.
\end{equation}
The swirling and magnetic field strength parameters make the differential Eq. (\ref{Hcon}) nonseparable, possibly resulting in the chaotic motion of particles. In the following, we explore the effects of the swirling parameter $j$ together with the magnetic field strength $B$, electric charge $Q$, magnetic charge $H$ and spin parameter $a$ on the motion of particles in the dyonic Kerr-Newman black hole immersed in the Melvin-swirling universe. 
	
\section{Chaotic motion of particles}

The reliable detection of chaotic behavior necessitates fast and accurate numerical methods. Because the motion of particles in chaotic regions exhibits the extreme sensitivity to initial conditions, tiny numerical errors could produce the pseudo-chaos. The corrected fifth-order Runge-Kutta method, proposed in Refs. \cite{DZMa1, DZMa2, Wu1}, incorporates velocity corrections ($\dot{r}$, $\dot{\theta}$) during integration and minimizes numerical deviations by pulling the solution back along a least-squares shortest path. Thus, we employ this method to solve the differential equations (\ref{wfE1})-(\ref{wfE3}) accurately in this study. For concreteness, we take the parameters $\{$$M=1$, $E=0.95$, $L=2.4M$$\}$ and set the initial conditions $\{$$ r(0)=10.5$, $\dot{r}(0)=0$,  $\theta(0)=\pi/2$$\}$, and consider different values of the swirling parameter $j$, magnetic field strength $B$, electric charge $Q$, magnetic charge $H$ and spin parameter $a$. In the Appendix, we still use the same values of the parameters $\{$$M$, $E$, $L$$\}$ and initial conditions to investigate the chaotic motion of particles in sub-cases of this dyonic Kerr-Newman-Melvin-swirling spacetime by removing conical singularities and removing both conical singularities and Dirac strings.

\subsection{Poincar\'{e} section}

\begin{figure}[htbp!]
\includegraphics[width=5.5cm ]{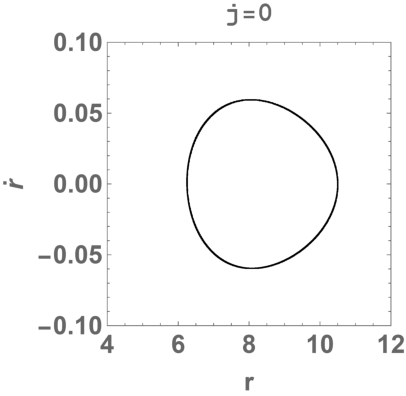}
\includegraphics[width=5.5cm ]{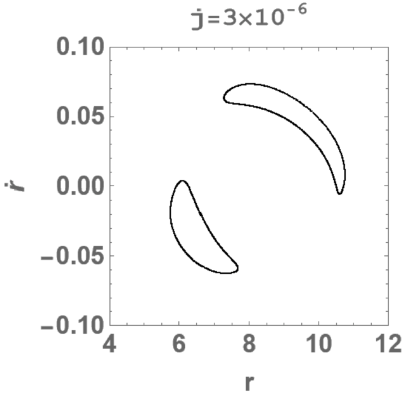}
\includegraphics[width=5.5cm ]{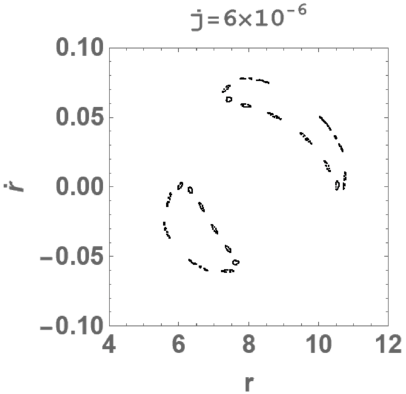}
\includegraphics[width=5.5cm ]{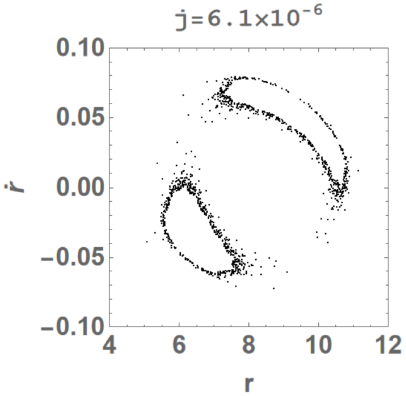}
\includegraphics[width=5.5cm ]{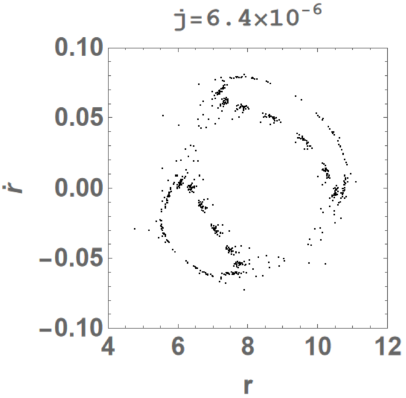}
\includegraphics[width=5.5cm ]{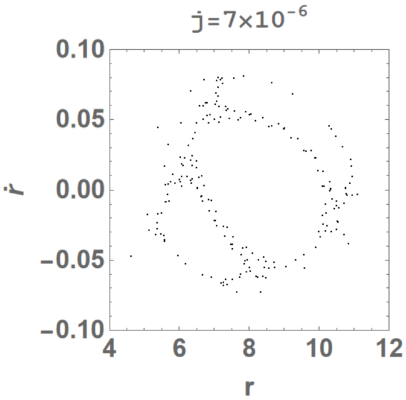}
\caption{The change of the Poincar\'{e} section ($\theta = \frac{\pi}{2}$) with the swirling parameter $j$ for the motion of a particle around the dyonic Kerr-Newman black hole immersed in the Melvin-swirling universe. Here we set $r(0) = 10.5$, $B = 10^{-4}$, $Q = 0.1$, $H = 0.1$, $a = 0.1$, $M = 1$, $E = 0.95$ and $L = 2.4M$. }\label{fig1}
\end{figure}

\begin{figure}[htbp!]
	\includegraphics[width=5.5cm ]{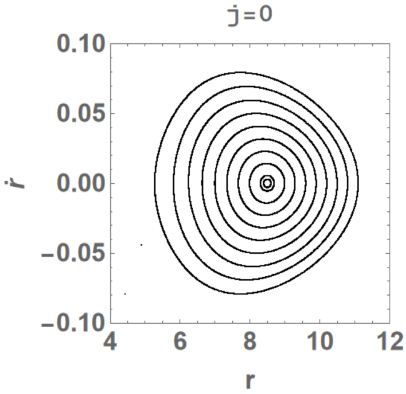}
	\includegraphics[width=5.5cm ]{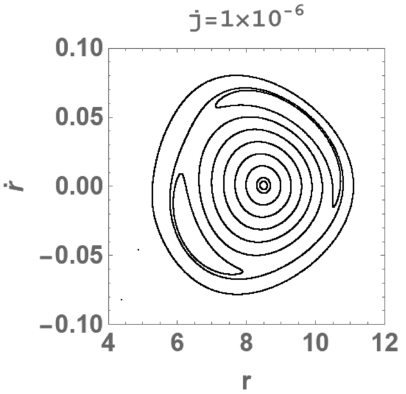}
	\includegraphics[width=5.5cm ]{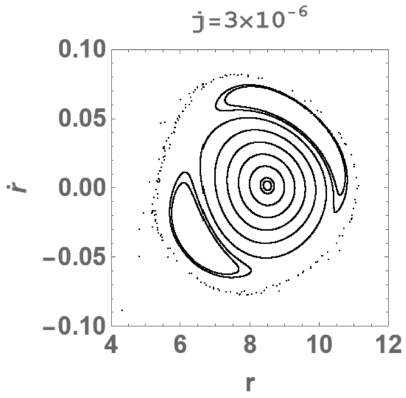}
	\includegraphics[width=5.5cm ]{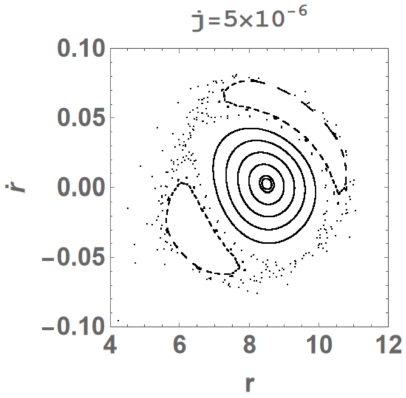}
	\includegraphics[width=5.5cm ]{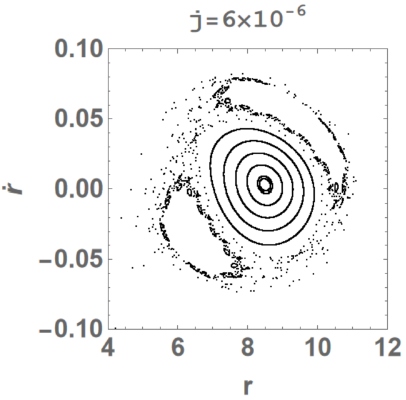}
	\includegraphics[width=5.5cm ]{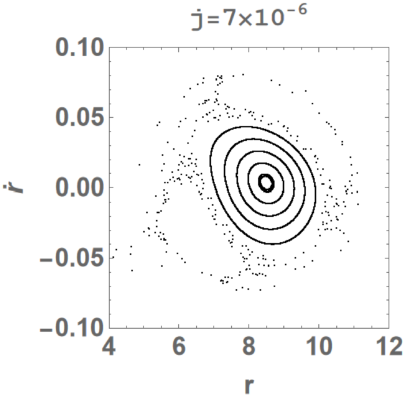}
	\includegraphics[width=5.5cm ]{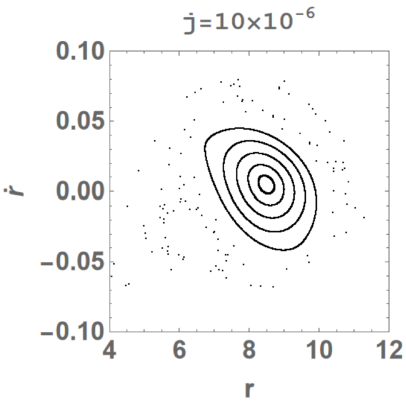}
	\includegraphics[width=5.5cm ]{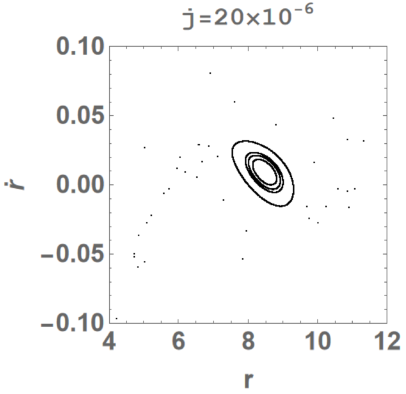}
	\includegraphics[width=5.5cm ]{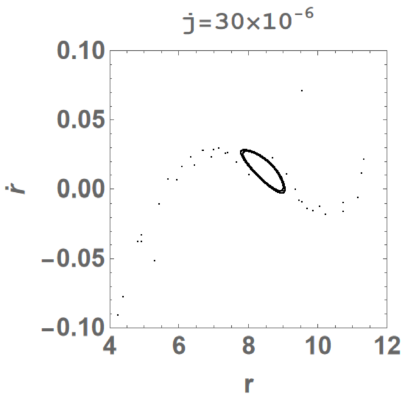}
	\caption{The change of the Poincar\'{e} section ($\theta = \frac{\pi}{2}$) with the swirling parameter $j$ for the motion of particles around the dyonic Kerr-Newman black hole immersed in the Melvin-swirling universe. Here we set $B = 10^{-4}$, $Q = 0.1$, $H = 0.1$, $a = 0.1$, $M = 1$, $E = 0.95$ and $L = 2.4M$.}\label{fig2}
\end{figure}

The Poincar\'{e} section, which can distinguish between regular and chaotic motion in a four-dimensional phase space, is defined as the intersection of the trajectory in a continuous dynamical system with a prescribed hypersurface that is transverse to the trajectory in the phase space. The distribution of points from orbits of the particle in the Poincar\'{e} section reflects the regularity or chaoticity of the motion. 

In Fig.~\ref{fig1}, we show the change of Poincar\'{e} sections $(r,\dot{r})$ for different swirling parameters $j$ in  the dyonic Kerr-Newman black hole immersed in the Melvin-swirling universe. We find that when $j \leq 6 \times 10^{-6}$, the motion
of the particle remains regular. Interestingly, for $j = 3 \times 10^{-6}$, an island chain of two secondary Kolmogorov-Arnold-Moser (KAM) tori appears, belonging to the same trajectory. Increasing the swirling parameter to $j \geq 6.1 \times 10^{-6}$, we observe that  the breakdown of the quasi-periodic KAM tori, and the Poincar\'{e} section exhibits scattered discrete points, indicating the chaotic motion. As $j \geq 7 \times 10^{-6}$, 
we note that the number of discrete points in the Poincar\'{e} section decreases. This is because the particle eventually either falls into the event horizon or escapes to spatial infinity. Obviously, the swirling parameter $j$ brings richer physics in the motion of particles.

\begin{figure}[htbp!]
\includegraphics[width=5.5cm ]{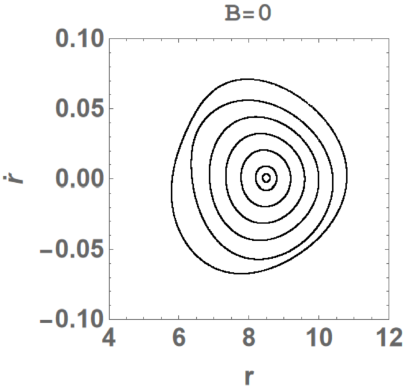}
\includegraphics[width=5.5cm ]{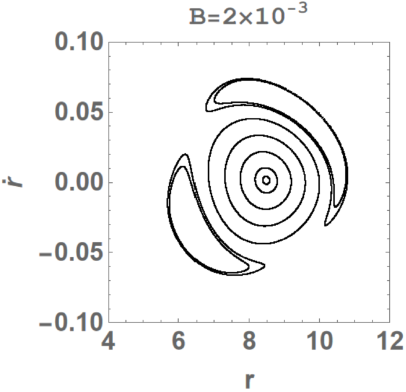}
\includegraphics[width=5.5cm ]{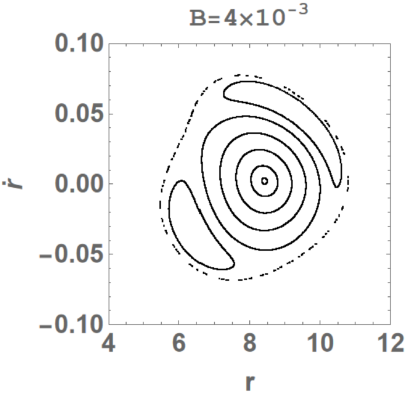}
\includegraphics[width=5.5cm ]{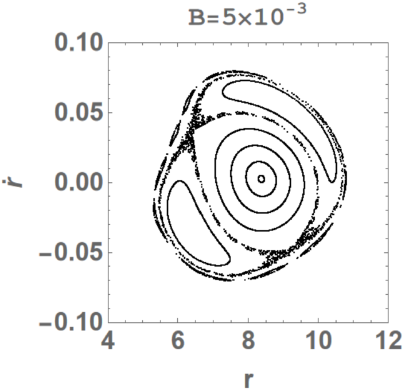}
\includegraphics[width=5.5cm ]{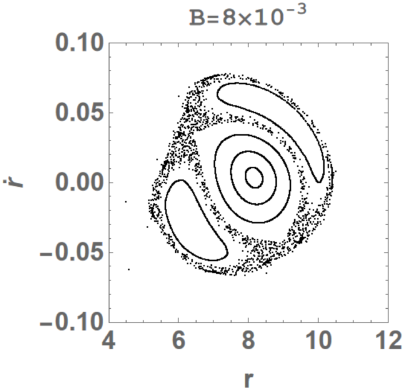}
\includegraphics[width=5.5cm ]{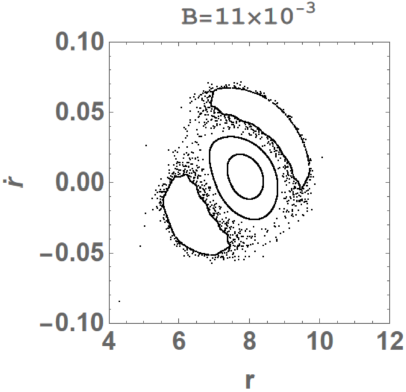}
\includegraphics[width=5.5cm ]{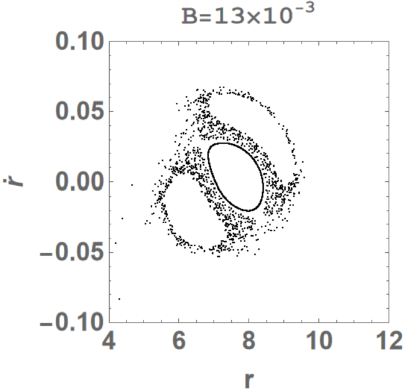}
\includegraphics[width=5.5cm ]{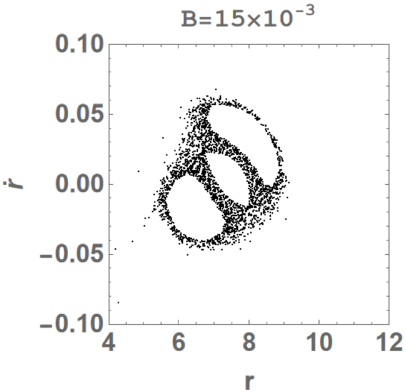}
\includegraphics[width=5.5cm ]{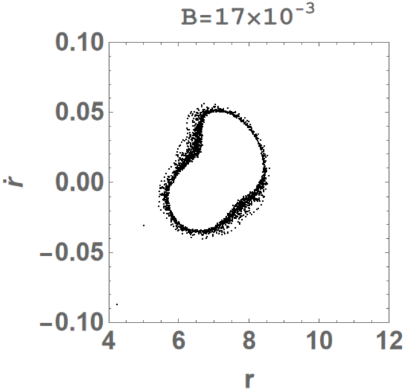}
\caption{The change of the Poincar\'{e} section ($\theta = \frac{\pi}{2}$) with the magnetic ﬁeld strength parameter $B$ for the motion of particles around the dyonic Kerr-Newman black hole immersed in the Melvin-swirling universe. Here we set $j = 10^{-6}$, $Q = 0.1$, $H = 0.1$, $a = 0.1$, $M = 1$, $E = 0.95$ and $L = 2.4M$.}\label{fig3}
\end{figure}

\begin{figure}[htbp!]
\includegraphics[width=4.1cm ]{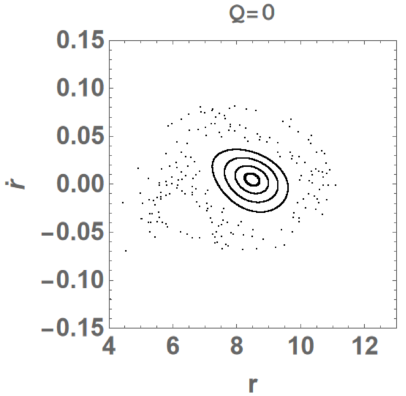}
\includegraphics[width=4.1cm ]{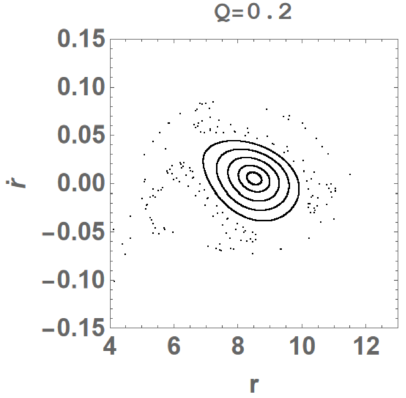}
\includegraphics[width=4.1cm ]{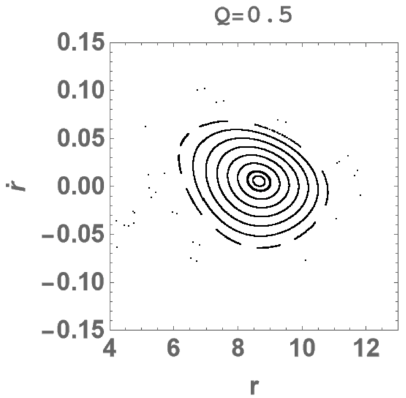}
\includegraphics[width=4.1cm ]{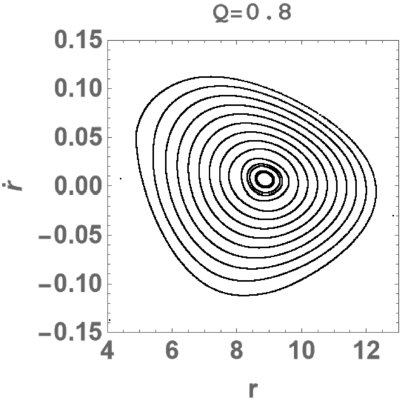}
\includegraphics[width=4.1cm ]{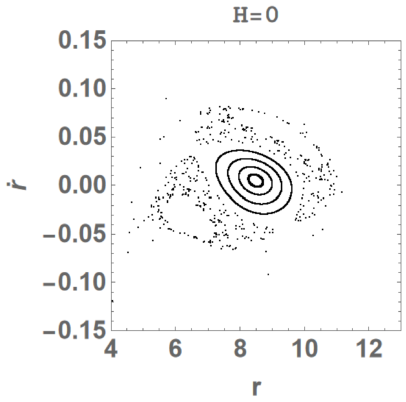}
\includegraphics[width=4.1cm ]{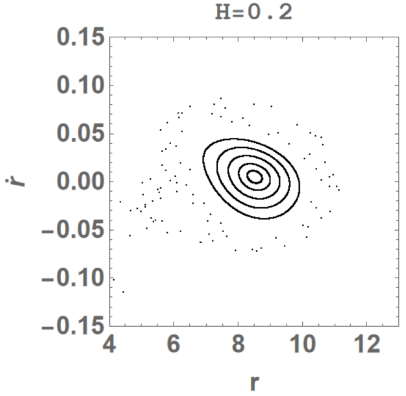}
\includegraphics[width=4.1cm ]{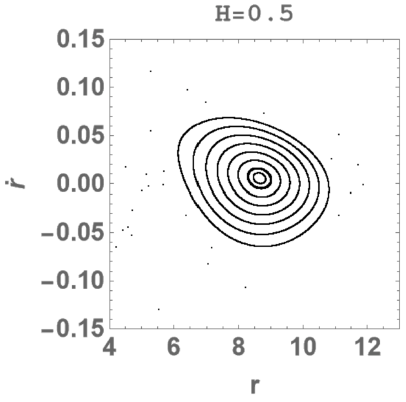}
\includegraphics[width=4.1cm ]{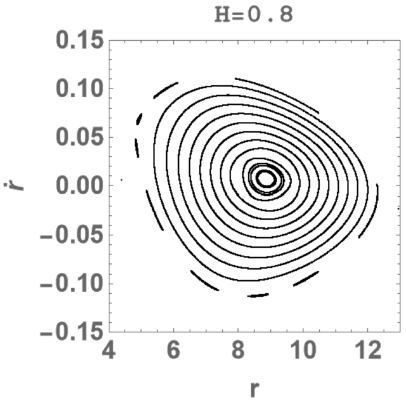}
\includegraphics[width=4.1cm ]{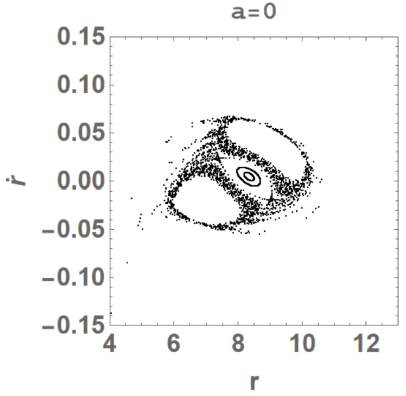}
\includegraphics[width=4.1cm ]{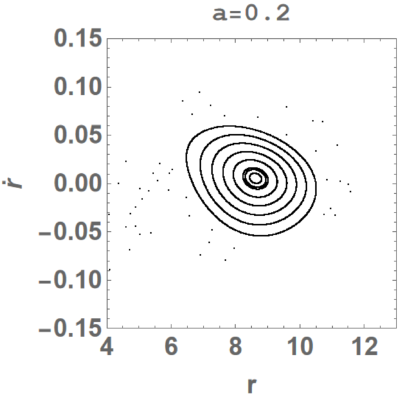}
\includegraphics[width=4.1cm ]{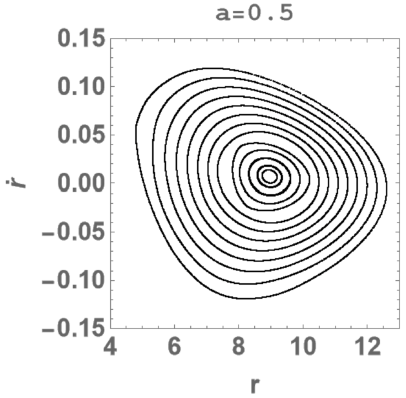}
\includegraphics[width=4.1cm ]{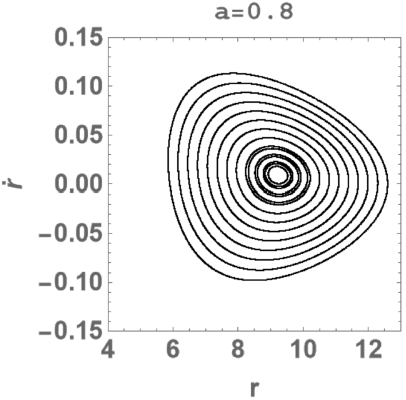}
\caption{The change of the Poincar\'{e} section ($\theta = \frac{\pi}{2}$) with the electric charge parameter $Q$ (top, $H = 0.1$ and $a = 0.1$), magnetic charge parameter $H$ (middle, $Q = 0.1$ and $a = 0.1$) and spin parameter $a$ (bottom, $Q = 0.1$ and $H = 0.1$) for the motion of particles around the dyonic Kerr-Newman black hole immersed in the Melvin-swirling universe. Here we set $j =10^{-5}$, $B =10^{-4}$, $M = 1$, $E = 0.95$ and $L = 2.4M$.}\label{fig4}
\end{figure}

Now we proceed to more motion orbits of particles. In Fig. \ref{fig2}, we plot the Poincar\'{e} sections with more motion orbits around the dyonic Kerr-Newman black hole immersed in the Melvin-swirling universe. We find that, as the swirling parameter $j$ increases, the number of regular motion orbits decreases, but the number of chaotic orbits and the chaotic region increase, which is similar to that in the dyonic Kerr black hole case \cite{Cao}. In Fig. \ref{fig3}, for the fixed swirling parameter $j=10^{-6}$, we present the Poincar\'{e} sections with different magnetic ﬁeld strength parameters $B$ around the dyonic Kerr-Newman black hole immersed in the Melvin-swirling universe. We observe that the number of chaotic orbits increases with the increase of the magnetic field strength parameter $B$, which is similar to the effect of the swirling parameter $j$ shown in Fig. \ref{fig2} and agrees well with the results in \cite{Yang}. Thus, we point out that the non-integrability of the motion of particles increases as the swirling parameter $j$ or magnetic field strength parameter $B$ increases.

In order to obtain the effects of the electric charge parameter $Q$, magnetic charge parameter $H$ and spin parameter $a$ on the motion of particles, we give the change of the Poincar\'{e} section with $Q$,  $H$ and $a$ for the dyonic Kerr-Newman black hole immersed in the Melvin-swirling universe in Fig. \ref{fig4}. We note that both the number of chaotic orbits and the chaotic region decrease as $Q$, $H$ or $a$ increases, which is different from the effects of the swirling parameter $j$ or magnetic field strength parameter $B$.

\subsection{Fast Lyapunov Indicator}

\begin{figure}[htbp!]
\includegraphics[width=5.3cm ]{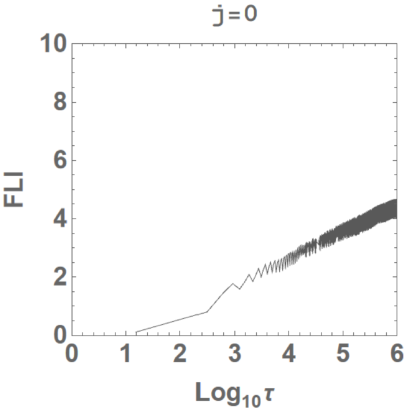}
\includegraphics[width=5.3cm ]{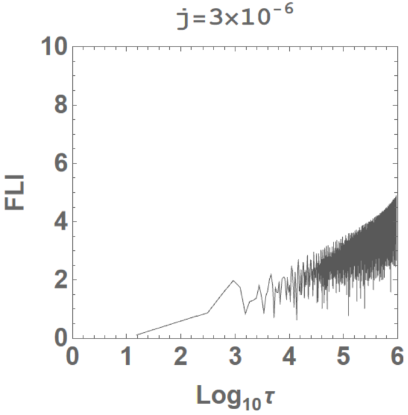}
\includegraphics[width=5.3cm ]{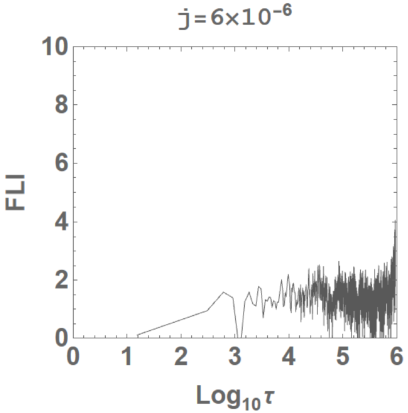}
\includegraphics[width=5.3cm ]{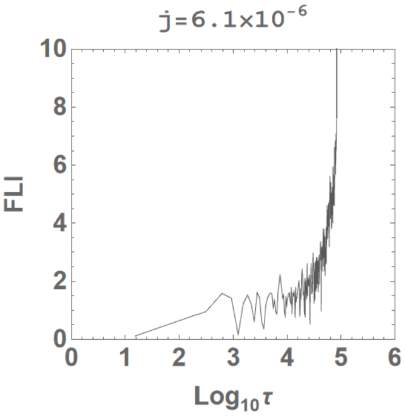}
\includegraphics[width=5.3cm ]{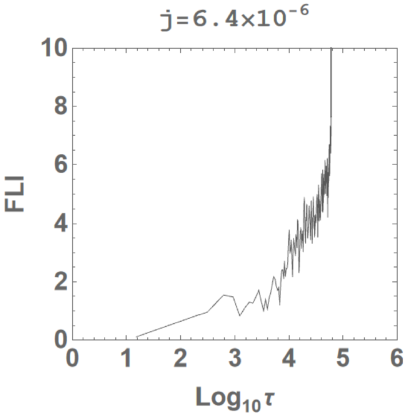}
\includegraphics[width=5.3cm ]{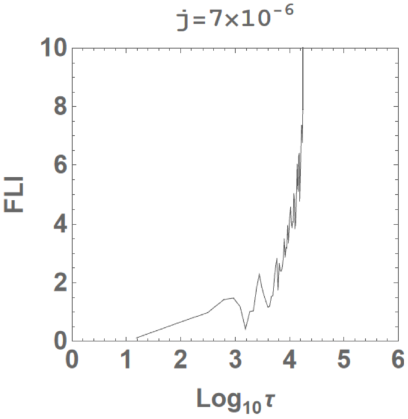}
\caption{The fast Lyapunov indicator (FLI) with the swirling parameter $j$ for the motion of particles shown in Fig. \ref{fig1}. }\label{fig5}
\end{figure}

FLI is a fast and effective tool for detecting chaotic behavior of particles \cite{froeschle1997,Wu2,Wu3}. In the curved spacetime, the FLI based on the two-particle method is defined as
\begin{equation}
FLI(\tau)=-k[1+\log_{10}d(0)]+\log_{10}\frac{|d(\tau)|}{|d(0)|},
\end{equation}
with $d(\tau)=\sqrt{|g_{\mu\nu}\Delta x^{\mu}\Delta x^{\nu}|}$. Here, $\Delta x^\mu$ is the deviation vector between two neighboring trajectories at proper time $\tau$.
To prevent numerical saturation caused by  the rapid separation of two adjacent trajectories, we introduce the renormalization number $k$.
Whenever $d(\tau) = 1$, the renormalization step is triggered: the value of $k$ is increased by one, and $d(\tau)$ is reset to its initial value $d(0)$. It should be noted that the $\text{FLI}(\tau)$ grows algebraically with time for the regular or periodic orbit, but grows exponentially for the chaotic orbit.

In Fig. \ref{fig5}, we plot the $\text{FLI}(\tau)$ with the different swirling parameters $j$ for the selected initial orbit presented in Fig. \ref{fig1}. As shown in Fig. \ref{fig5}, the change of $\text{FLI}(\tau)$ over an integration time of $10^{6}$ shows that, for $j \geq 6.1 \times 10^{-6}$, the $\text{FLI}(\tau)$ exhibits the exponential growth with the increase of $\tau$, indicating the orbits are chaotic. In contrast, for $j \leq 6 \times 10^{-6}$, the $\text{FLI}(\tau)$ grows linearly with $\tau$, which is characteristic of regular cases. The results are consistent with those obtained by the Poincar\'{e} section given in Fig. \ref{fig1}.
	
\subsection{Recurrence analysis}

\begin{figure}[htbp!]
	\includegraphics[width=5.5cm ]{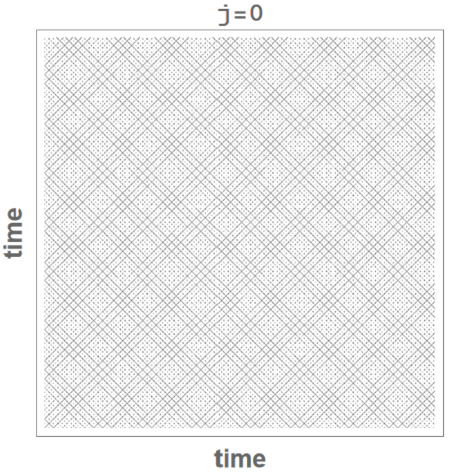}
	\includegraphics[width=5.5cm ]{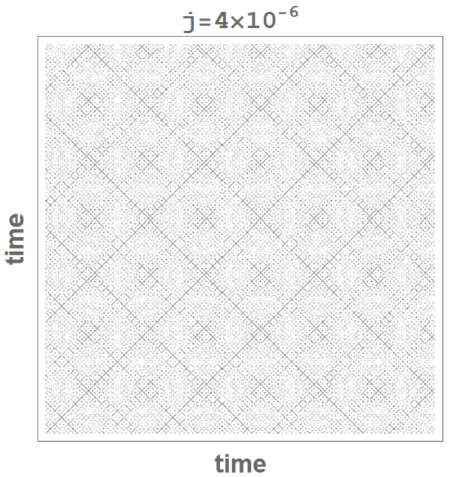}
	\includegraphics[width=5.5cm ]{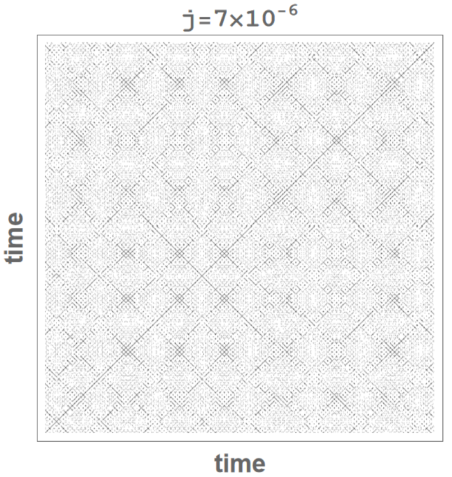}
	\caption{The change of the recurrence plots with the swirling parameter $j$ for the motion of particles around the dyonic Kerr-Newman black hole immersed in the Melvin-swirling universe. Here we set $r(0) = 10.5$, $B = 10^{-4}$, $Q = 0.1$, $H = 0.1$, $a = 0.1$, $M = 1$, $E = 0.95$ and $L = 2.4M$. }\label{fig6}
\end{figure}

The recurrence analysis \cite{Eckmann,Marwan,Kopacek:2010yr,Kopacek:2010at,Kovar:2013pf} is a novel approach for distinguishing the chaos from the regular motion in relativistic systems, and the recurrence plots visualize the recurrences of an orbit into the vicinity of previously reached phase-space points. This method examines the binary values constructed from the phase space trajectory $\mathbf{x}(\tau)$ by evaluating the recurrence matrix $\mathbf{R}_{ij}$, defined as
\begin{equation}
	\label{rpdef}
	\mathbf{R}_{ij}(\varepsilon)=\Theta(\varepsilon-||\textbf{x}(i)-\textbf{x}(j)||), ~~
	i,j=1,...,N,
\end{equation}
where $\varepsilon=k\sigma$ is the pre-defined threshold parameter with a proportionality constant $k$ and the standard mean deviation of the given data set $\sigma$. The Heaviside function $\Theta(\vartheta)=0$ for $\vartheta<0$ and $1$ for $\vartheta\geq0$. Here, $N$ stands for the sampling frequency which is applied to the examined time period of the trajectory $\mathbf{x}(\tau)$, and the space norm $||\;.\;||$ is the Euclidean norm. In the recurrence plots, a black dot represents $\mathbf{R}_{ij}=1$ and a white dot corresponds to $\mathbf{R}_{ij}=0$. Both axes denote a time segment over which the data set, i.e., the phase-space vector, is being examined. The presence of numerous diagonal lines parallel to the main diagonal implies the regular motion, whereas short, disrupted diagonals or the absence of such lines suggests the chaos. 

In Fig.~\ref{fig6}, the recurrence plots for $j=0$ and $4 \times 10^{-6}$ exhibit diagonal line structures, corresponding to regular orbits. However, the recurrence plot for $j=7 \times 10^{-6}$ lacks complete diagonal structures, indicating the chaotic motion. 
This means that the larger swirling parameter $j$ makes it easier for the emergence of chaotic behavior in the motion of particles. Obviously, the recurrence analysis provides the cross-validation for our primary investigation of the motion of particles around the dyonic Kerr-Newman black hole immersed in the Melvin-swirling universe by using the Poincaré sections and FLI.

\subsection{Bifurcation diagram}

\begin{figure}[htbp!]
	\includegraphics[width=5.4cm ]{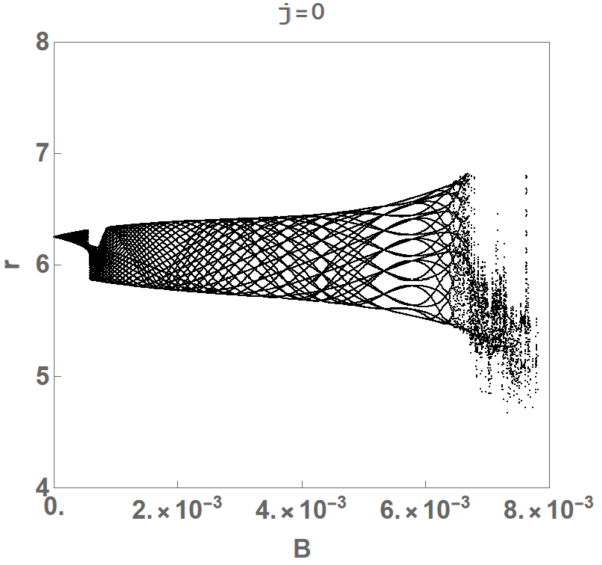}
	\includegraphics[width=5.4cm ]{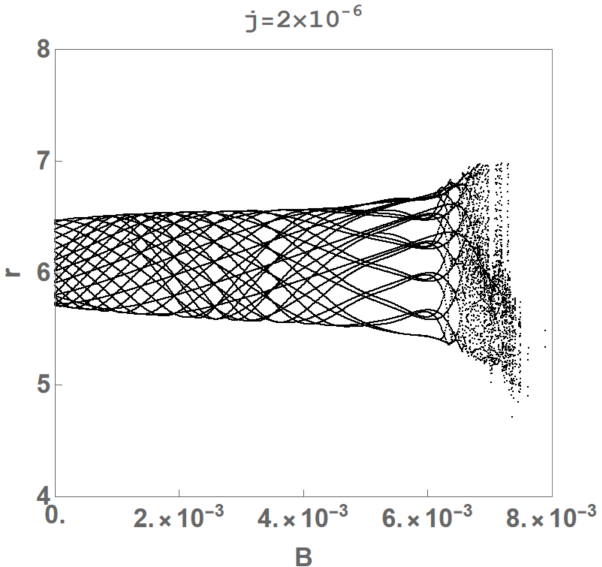}
	\includegraphics[width=5.4cm ]{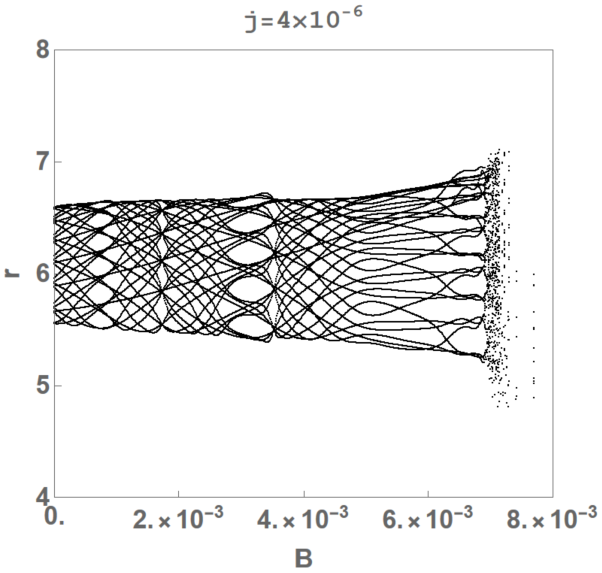}
	\includegraphics[width=5.4cm ]{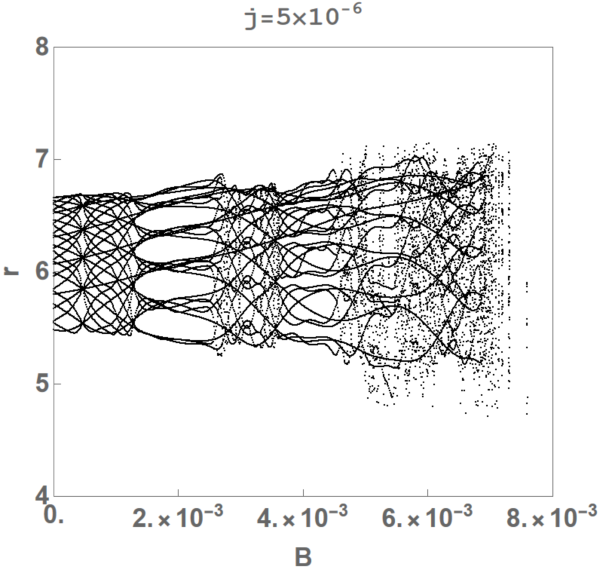}
	\includegraphics[width=5.4cm ]{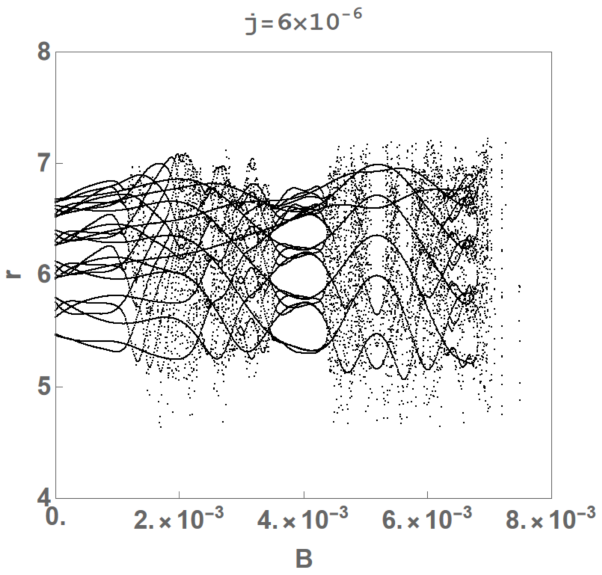}
	\includegraphics[width=5.4cm ]{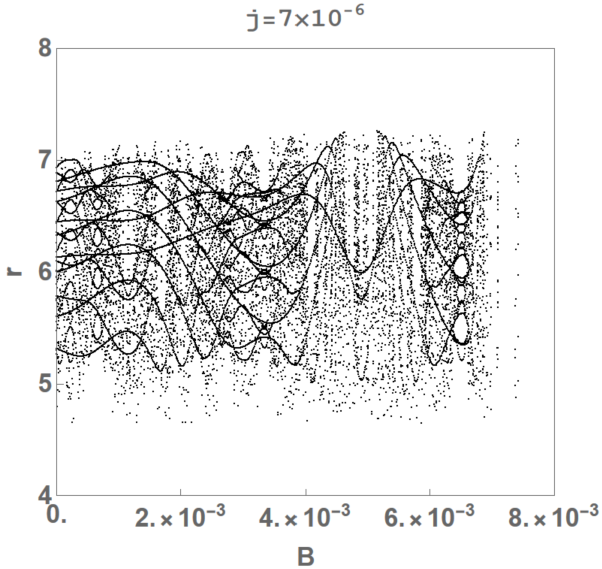}
	\caption{The change of the bifurcation with the magnetic ﬁeld strength parameter $B$ for different values of the swirling parameter $j$ in the spacetime of the dyonic Kerr-Newman black hole immersed in the Melvin-swirling universe. Here we set $Q =0.1$, $H = 0.1$, $a = 0.1$, $M = 1$, $E = 0.95$ and $L = 2.4M$. }\label{fig10}
\end{figure}

\begin{figure}[htbp!]
\includegraphics[width=5.4cm ]{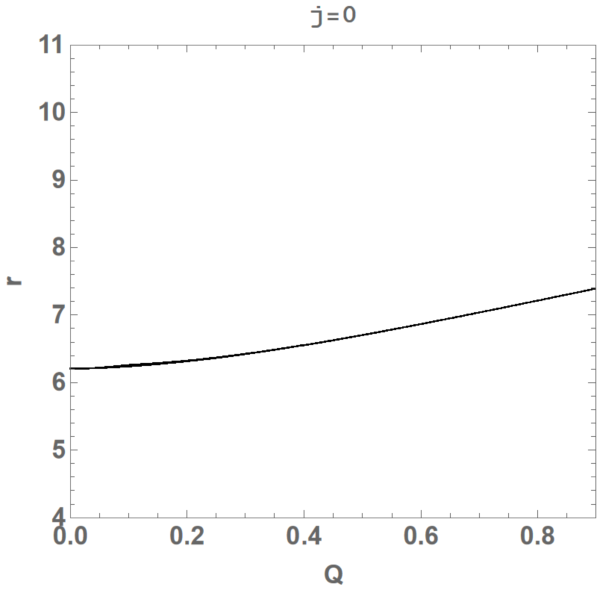}
\includegraphics[width=5.4cm ]{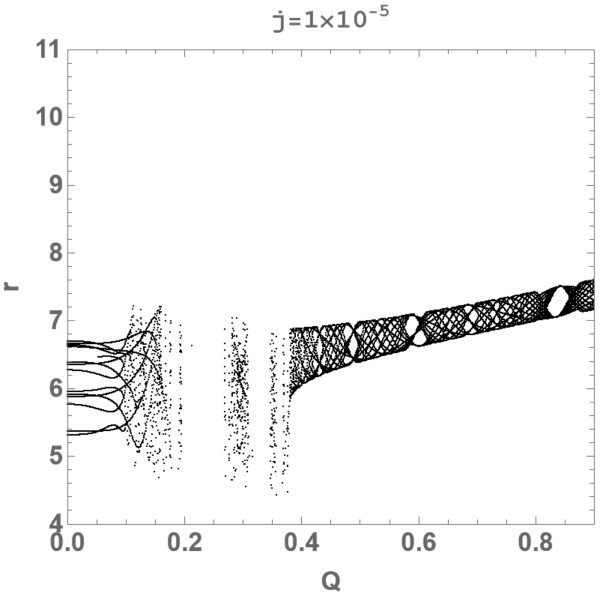}
\includegraphics[width=5.4cm ]{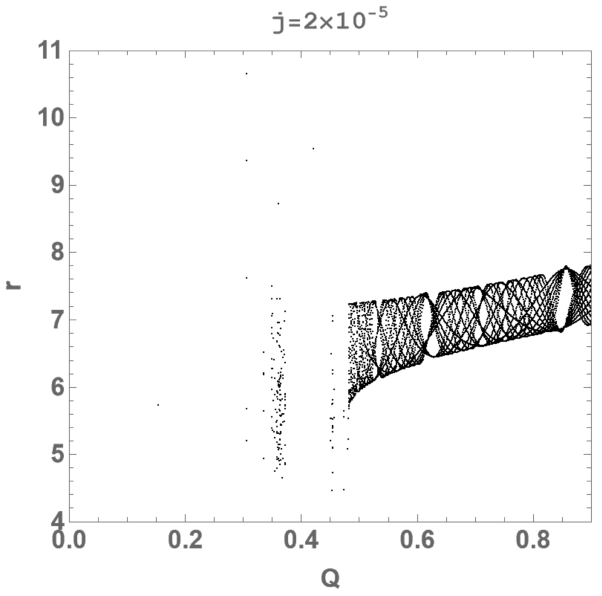}
\includegraphics[width=5.4cm ]{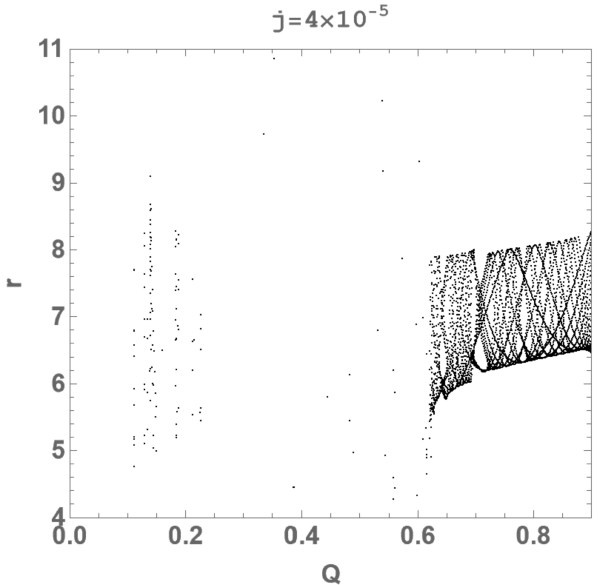}
\includegraphics[width=5.4cm ]{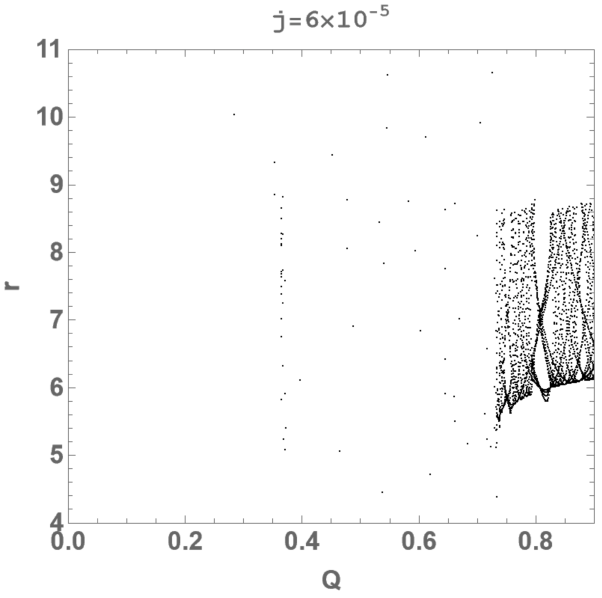}
\includegraphics[width=5.4cm ]{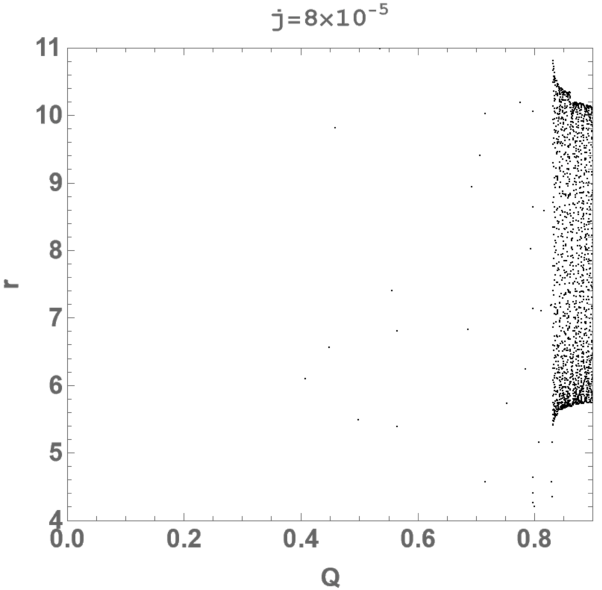}
\caption{The change of the bifurcation with the electric charge parameter $Q$ for different values of the swirling parameter $j$ in the spacetime of the dyonic Kerr-Newman black hole immersed in the Melvin-swirling universe. Here we set $B =10^{-4}$, $H = 0.1$, $a = 0.1$, $M = 1$, $E = 0.95$ and $L = 2.4M$.}\label{fig7}
\end{figure}

\begin{figure}[htbp!]
\includegraphics[width=5.4cm ]{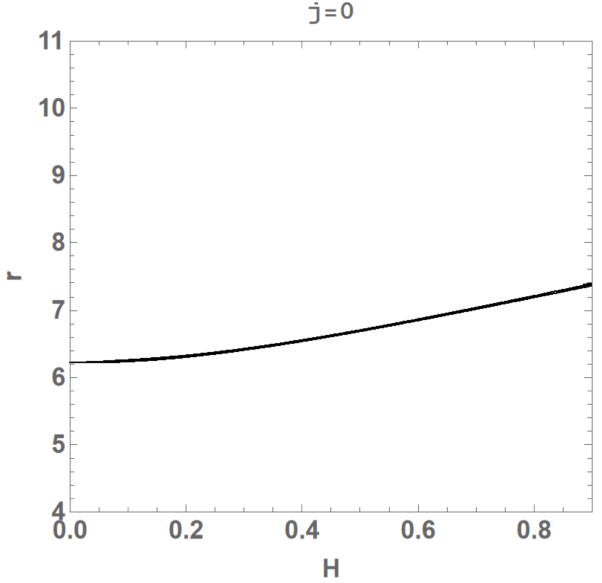}
\includegraphics[width=5.4cm ]{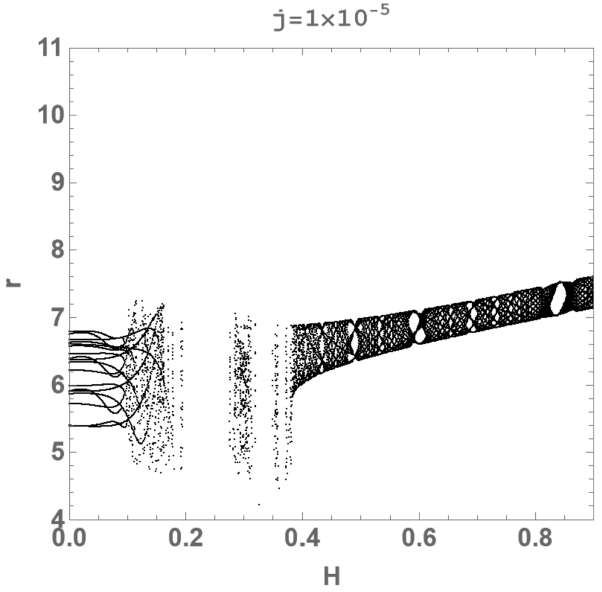}
\includegraphics[width=5.4cm ]{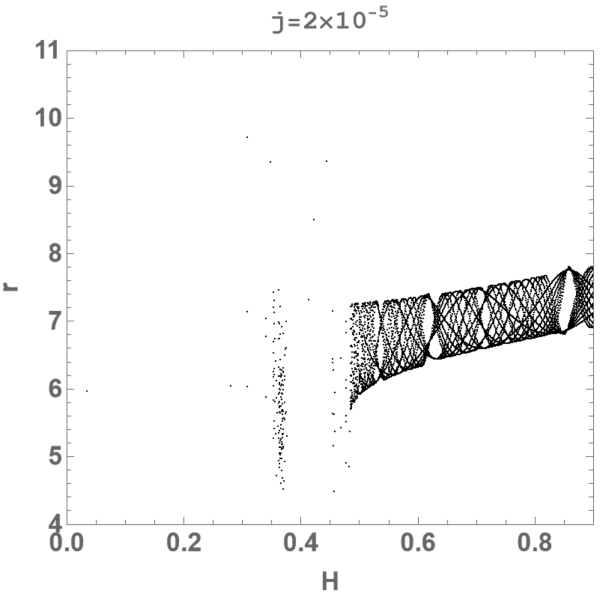}
\includegraphics[width=5.4cm ]{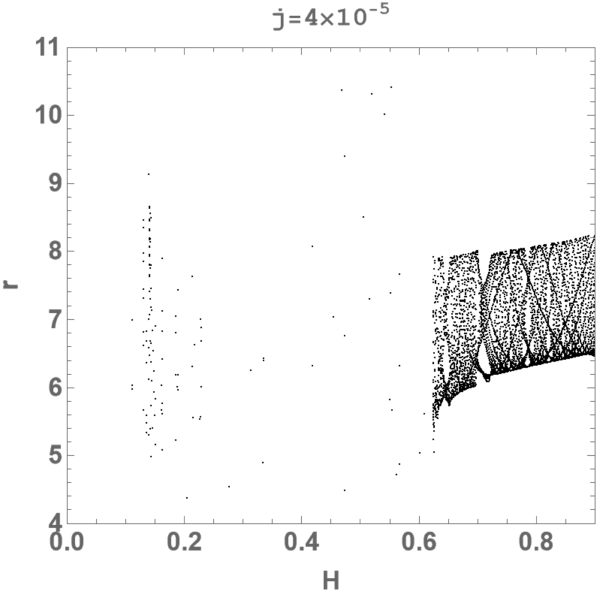}
\includegraphics[width=5.4cm ]{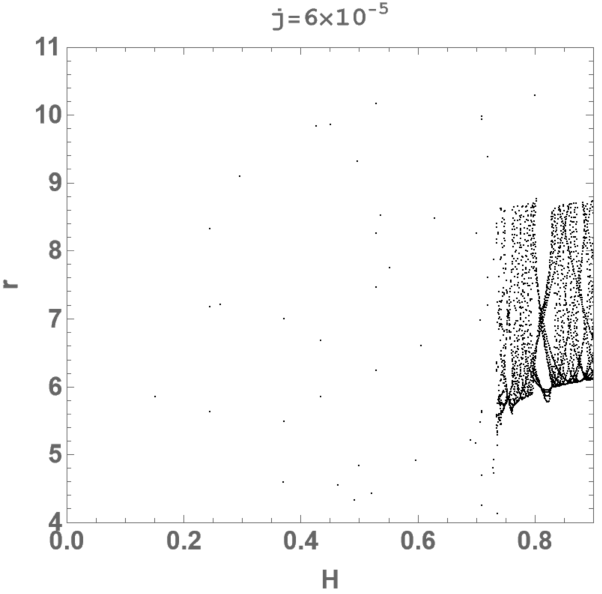}
\includegraphics[width=5.4cm ]{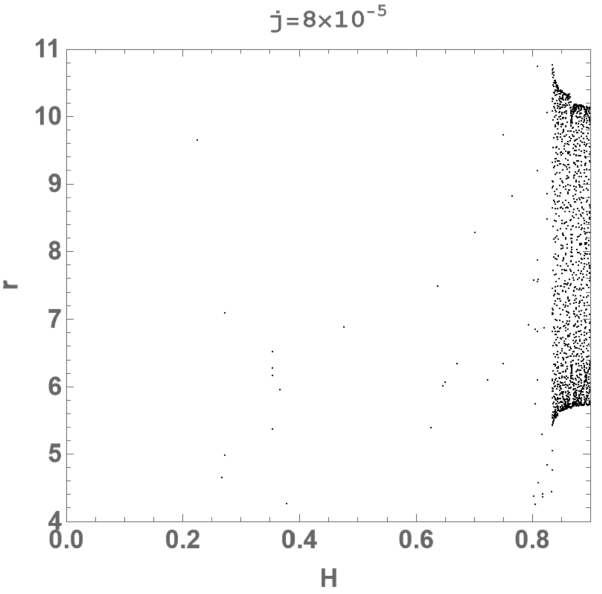}
\caption{The change of the bifurcation with the magnetic charge parameter $H$ for different values of the swirling parameter $j$ in the spacetime of the dyonic Kerr-Newman black hole immersed in the Melvin-swirling universe. Here we set $B =10^{-4}$, $Q = 0.1$, $a = 0.1$, $M = 1$, $E = 0.95$ and $L = 2.4M$. }\label{fig8}
\end{figure}

\begin{figure}[htbp!]
\includegraphics[width=5.4cm ]{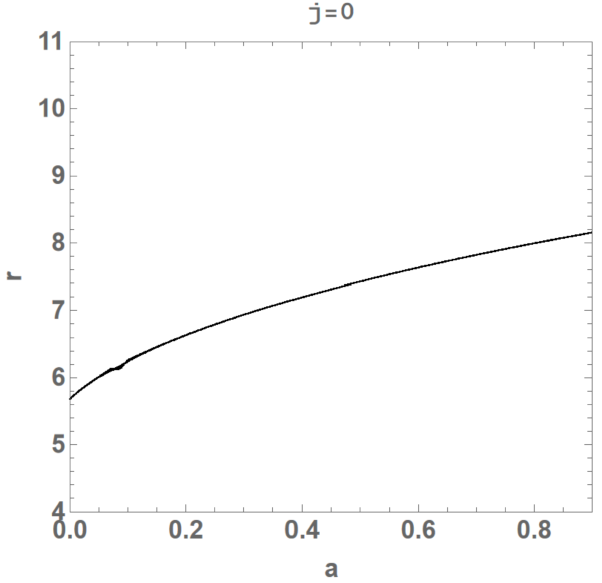}
\includegraphics[width=5.4cm ]{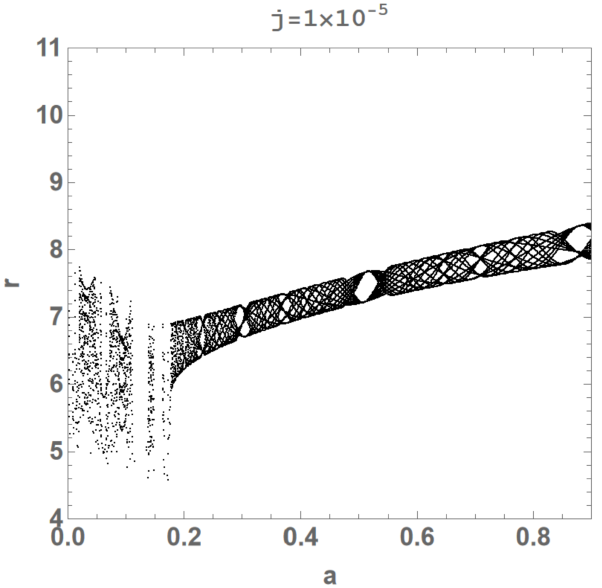}
\includegraphics[width=5.4cm ]{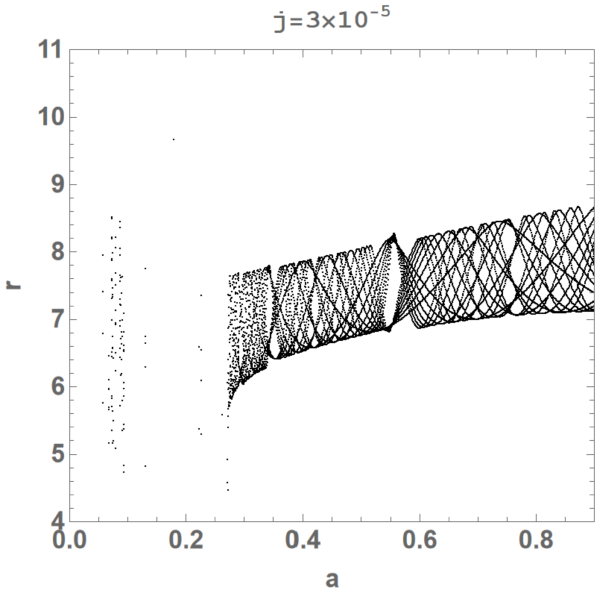}
\includegraphics[width=5.4cm ]{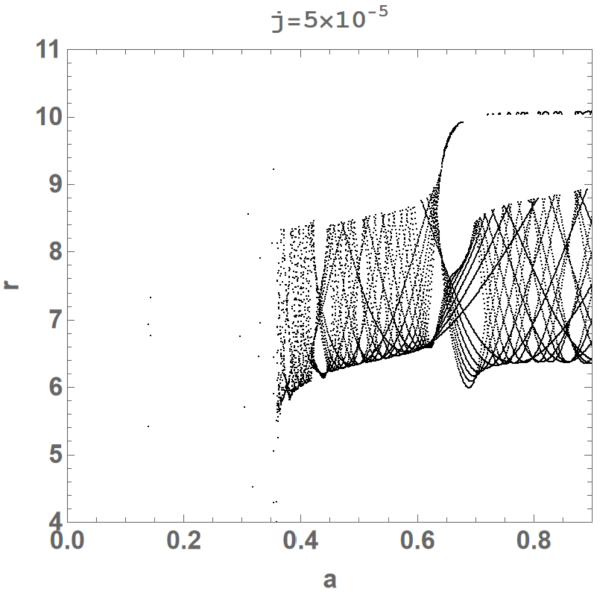}
\includegraphics[width=5.4cm ]{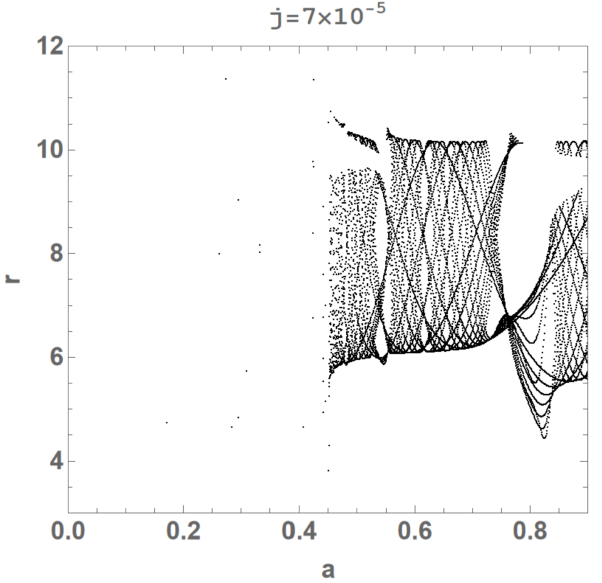}
\includegraphics[width=5.4cm ]{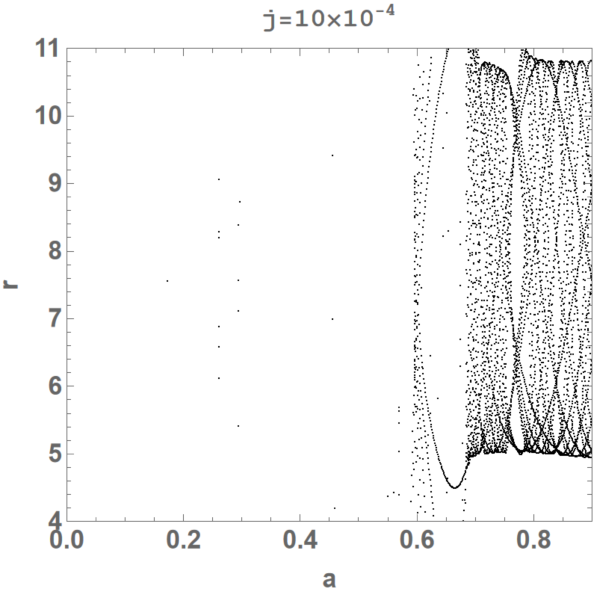}
\caption{The change of the bifurcation with the spin parameter $a$ for different values of the swirling parameter $j$ in the spacetime of the dyonic Kerr-Newman black hole immersed in the Melvin-swirling universe. Here we set $B =10^{-4}$, $Q = 0.1$, $H = 0.1$, $M = 1$, $E = 0.95$ and $L = 2.4M$. }\label{fig9}
\end{figure}

To further illustrate how the system's dynamical behavior depends on the black hole parameters, we can count on the bifurcation diagram. We present the change of the bifurcation with the magnetic ﬁeld strength parameter $B$ for different values of the swirling parameter $j$ in Fig. \ref{fig10} in order to get the effects of $B$ on the bifurcation diagram. We see that with the increase of $j$, the range of $B$ where the chaos occurs first decreases and then increases, and the corresponding lower limit of $B$ decreases. Therefore, we conclude that due to the presence of the swirling parameter $j$, the dynamical behavior of particles in the spacetime of a dyonic Kerr-Newman black hole immersed in the Melvin-swirling universe becomes much richer.

In Figs. \ref{fig7} - \ref{fig9}, we present the bifurcation diagrams of the radial coordinate $r(\tau)$ of particles with the electric charge parameter $Q$, magnetic charge parameter $H$ and spin parameter $a$ for different values of the swirling parameter $j$ in the spacetime of the dyonic Kerr-Newman black hole immersed in the Melvin-swirling universe. When $j = 0$ and $B = 10^{-4}$, the radial coordinate $r(\tau)$ is a periodic function and there is no bifurcation for the dynamical system, as shown in the first panel in Figs. \ref{fig7} - \ref{fig9}. Our results suggest that the motions of particles are regular in these cases, but the equations of motion are not separable due to the presence of the external magnetic field $B$. As the swirling parameter $j$ increases, the system exhibits periodic, chaotic and escaped solutions depending on the black hole parameters. Figs.~\ref{fig7} - \ref{fig9} also show that the lower bounds of the black hole parameters for chaotic orbits increase, indicating that the presence of $j$ changes the ranges of $Q$, $H$ and $a$ where the chaotic motion appears for particles. The influence of $j$ on the chaotic region of $a$ is in agreement with the findings reported in \cite{Cao}.

\subsection{Basins of attraction}

\begin{figure}[htbp!]
\includegraphics[width=6.68cm]{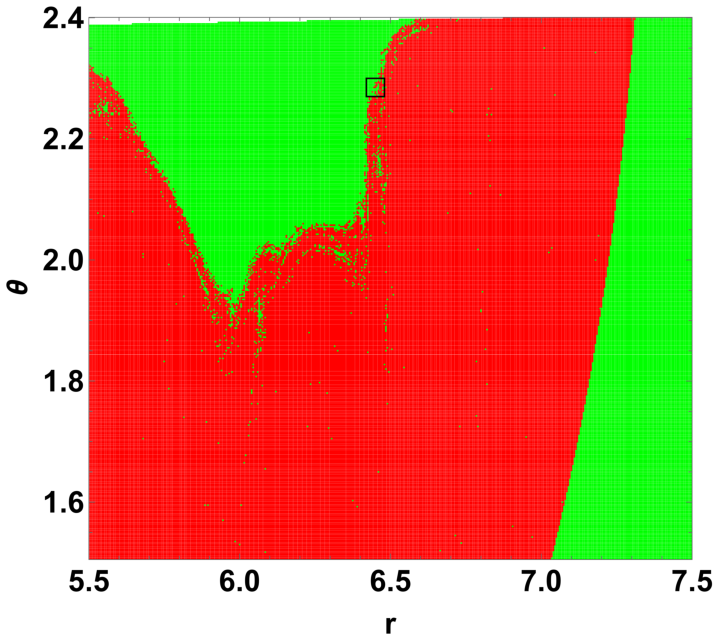}
\includegraphics[width=6.8cm]{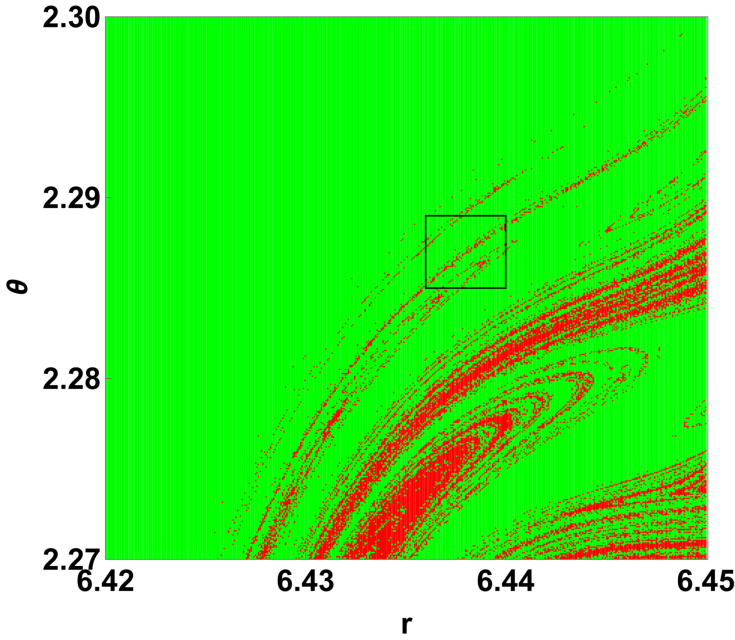}
\includegraphics[width=7.0cm]{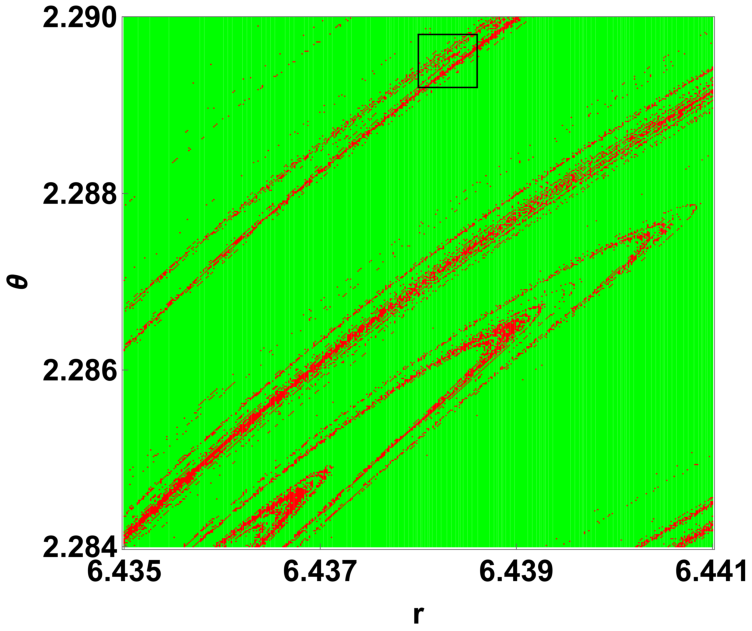}
\includegraphics[width=7.0cm]{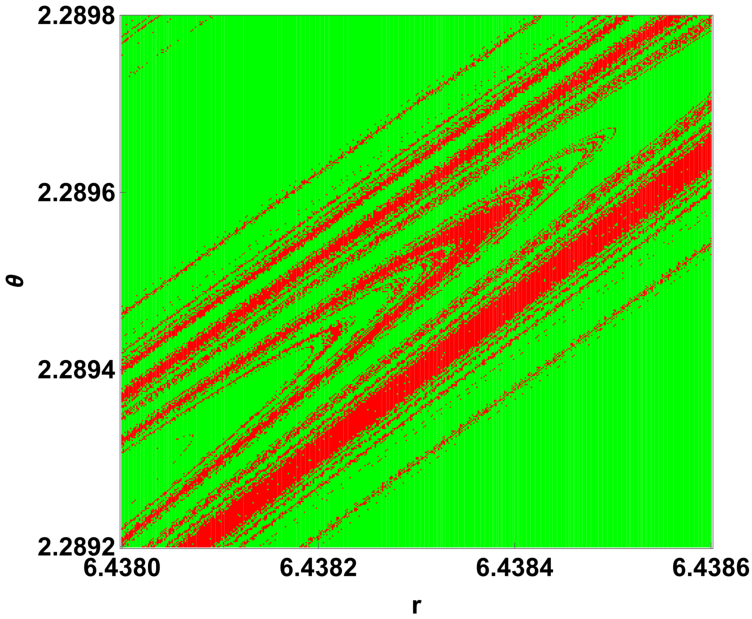}
\caption{The fractal basins of attraction for the motion of particles around the dyonic Kerr-Newman black hole immersed in the Melvin-swirling universe. Here we set $j =10^{-5}$, $B =10^{-4}$, $Q = 0.1$, $H = 0.1$, $a = 0.1$, $M = 1$, $E = 0.95$ and $L = 2.4M$. }\label{fig11}
\end{figure}

The basin boundaries of attractors can offer a signature of chaos \cite{DettmannFC,FrolovL,ZayasT,McDonald}. If the boundaries are fractal, the dynamics is chaotic. In Fig. \ref{fig11}, we plot the basins of attraction in a large subset of phase space for particles in the spacetime of the dyonic Kerr-Newman black hole immersed in the Melvin-swirling universe, where the initial conditions correspond to the points in these panels, $\dot{r} = 0$ and $\dot{\theta}$ given by the constraint (\ref{Hcon}), i.e., $h = 0$. The red points denote the particles that fall into the black hole, the blue points indicate the particles that radially escape to spatial infinity, while the green points correspond to particles that oscillate around the black hole. For concreteness, the condition for the particle capture is given by $r \leq r_{+}$, the condition for the escape is set to be $r \geq 100r_{+}$, and the green points correspond to trajectories that are neither captured nor escaped after 100,000 iterations. As shown in Fig.~\ref{fig11}, the basin boundaries of the attractors exhibit self-similar fractal structures, indicating the presence of chaotic motion for particles around the dyonic Kerr-Newman black hole immersed in the Melvin-swirling universe.

\section{Summary}
	
In this work, we have employed several tools including the Poincar\'{e} section, fast Lyapunov indicator, recurrence analysis, bifurcation diagram and basins of attraction to investigate the motion of particles around a dyonic Kerr-Newman black hole immersed in the Melvin-swirling universe, which is a new exact solution of Einstein-Maxwell field equations which
represents a rotating black hole with both electric and magnetic charges immersed in a
universe which itself is also rotating and magnetized \cite{DiPinto}. We noticed that the swirling parameter $j$ and magnetic field strength $B$ make the equations of motion for particles nonseparable, and confirmed the presence of chaotic behavior in the motion in this dyonic Kerr-Newman-Melvin-swirling spacetime and its sub-cases by removing the conical singularities and removing both the conical singularities and the Dirac strings. We observed that the combination of the swirling parameter $j$ with the magnetic field strength $B$, electric charge $Q$, magnetic charge $H$ and spin parameter $a$ provides richer physics in the chaotic motion for particles.
We found that both the number of chaotic orbits and the chaotic region in the Poincar\'{e} section increase with the increase of swirling parameter $j$ or magnetic field strength $B$, which indicates that the non-integrability of the motion of particles increases as $j$ or $B$ increases. However, the electric charge $Q$, magnetic charge $H$ and spin parameter $a$ have different effects on the chaotic motion for particles, i.e., both the number of chaotic orbits and the chaotic region decrease as $Q$, $H$ or $a$ increases. Moreover, the bifurcation diagrams for chaotic region show that as the swirling parameter $j$ increases, the lower bounds of the black hole parameters $Q$, $H$, and $a$ for chaotic regions increase but the lower bound of $B$ decreases, which suggests that the presence of $j$ changes the ranges of $B$, $Q$, $H$ and $a$ where the chaotic motion appears for particles. Therefore, the swirling parameter $j$ enriches the dynamical behaviors of particles in the spacetime of the dyonic Kerr-Newman black hole immersed in the Melvin-swirling universe.

\begin{acknowledgments}

This work was supported by the National Natural Science Foundation of China (Grant Nos. 12275079, 12275078 and 12035005), National Key Research and Development Program of China (Grant No. 2020YFC2201400) and the innovative research group of Hunan Province (Grant No. 2024JJ1006).

\end{acknowledgments}
	
\appendix

\section{Chaotic motion of particles for the solution without conical singularities}
\label{kerreqs}

For the novel solution (\ref{metric}), one can find the singularities, i.e. conical singularities, Dirac and Misner strings, and curvature singularities. In Ref.  \cite{DiPinto}, Di Pinto \emph{et al.} gave the solution without conical singularities from the spacetime (\ref{metric}) by choosing an appropriate value for the swirling parameter
\begin{equation}
j = - \frac{ B H \bigl( 4 + B^2 Z^2 \bigr)}{16 a M - 4 B Q Z^2},
\end{equation}
and by rescaling the azimuthal angle as $\phi \mapsto \frac{2\pi}{\delta \phi} \phi$ while maintaining the usual periodicity of $2\pi$
\begin{equation}
	\delta\phi = \frac{32 \pi \bigl( B Q Z^2 - 4 a M \bigr)^2\Bigl( 16 a^2 M^2 - 8 a B M Q Z^2 + B^2 Z^6\Bigr)^{-1}}{16 + 8 B^2 \bigl( H^2 + 3 Q^2 \bigr) + 32 a B^3 M Q + B^4 \bigl(16 a^2 M^2 + Z^4 \bigr)}.
\end{equation}
Since our focus is the swirling universe, we will keep the swirling parameter $j$ instead of the charge $H$ in the following analysis.

\begin{figure}[htbp!]
	\includegraphics[width=4.1cm ]{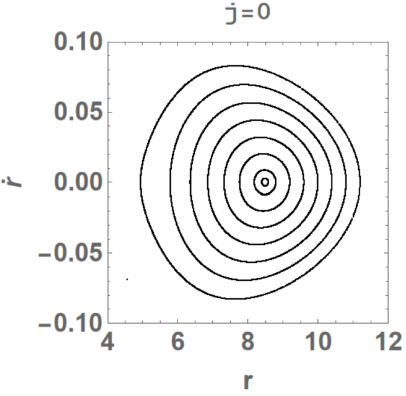}
	\includegraphics[width=4.1cm ]{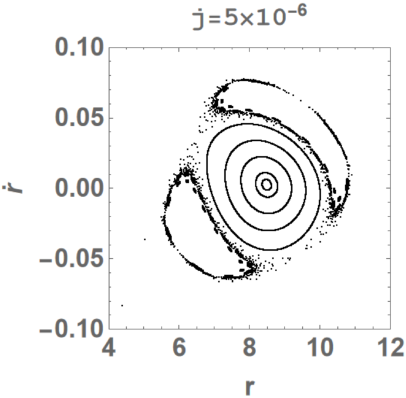}
	\includegraphics[width=4.1cm ]{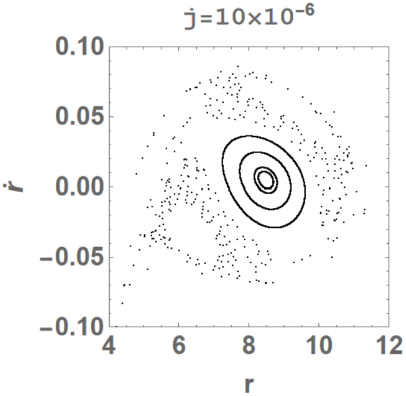}
	\includegraphics[width=4.1cm ]{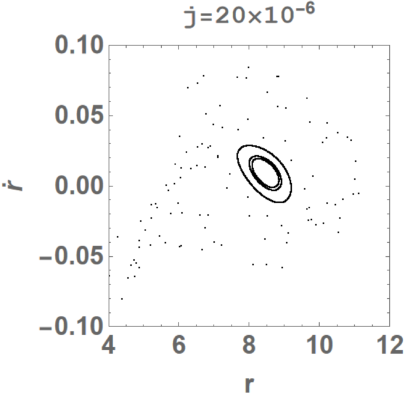}
	\includegraphics[width=4.1cm ]{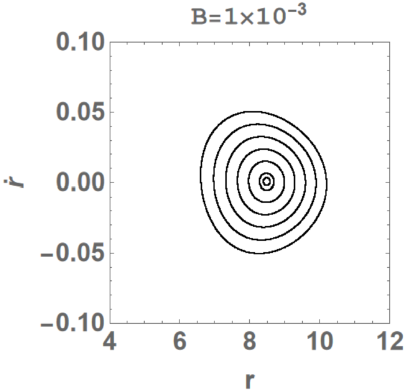}
	\includegraphics[width=4.1cm ]{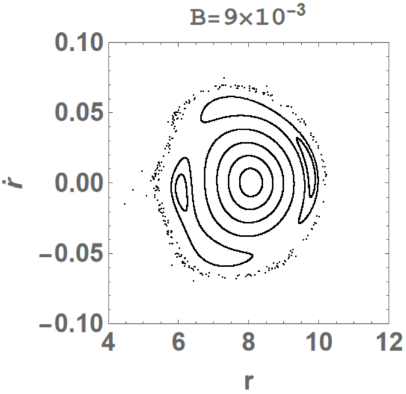}
	\includegraphics[width=4.1cm ]{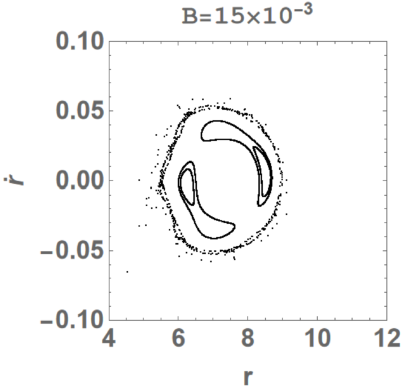}
	\includegraphics[width=4.1cm ]{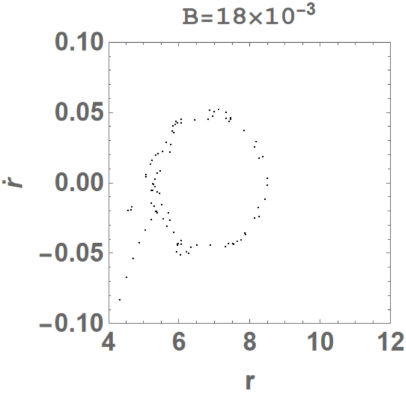}
	\includegraphics[width=4.1cm ]{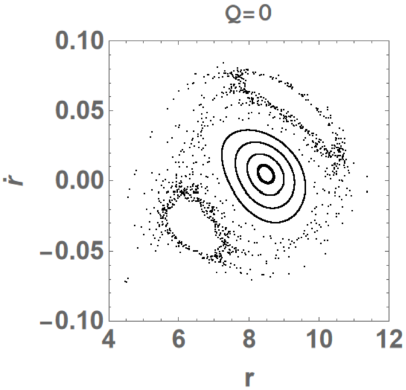}
	\includegraphics[width=4.1cm ]{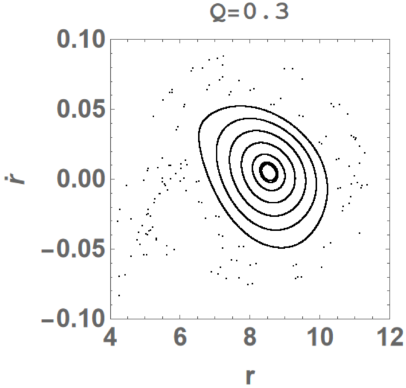}
	\includegraphics[width=4.1cm ]{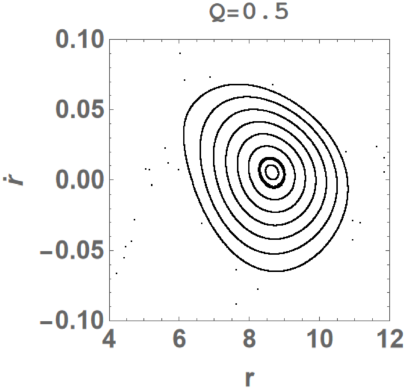}
	\includegraphics[width=4.1cm ]{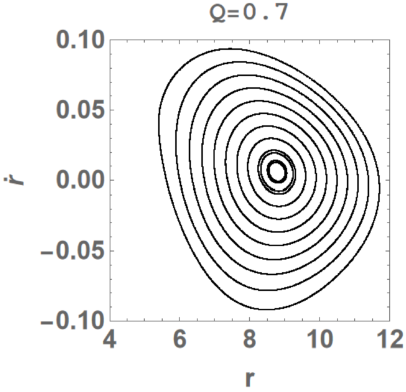}
	\includegraphics[width=4.1cm ]{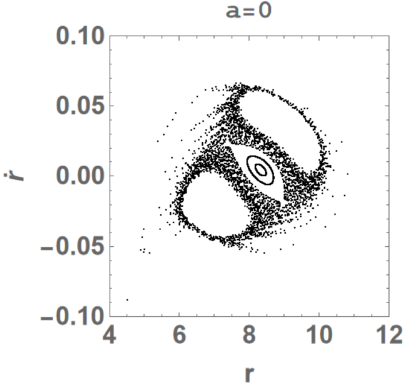}
	\includegraphics[width=4.1cm ]{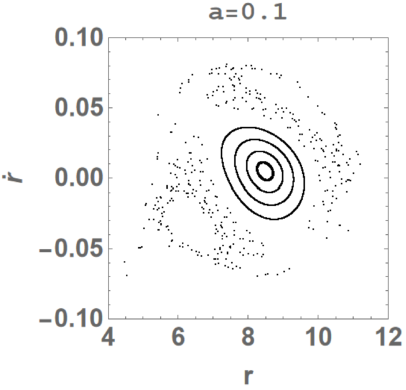}
	\includegraphics[width=4.1cm ]{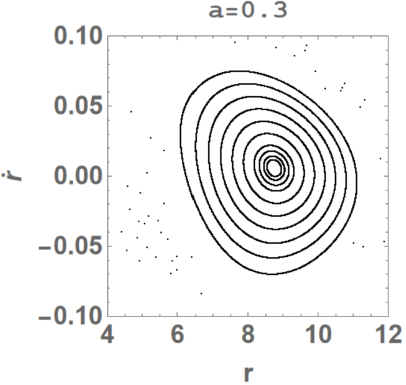}
	\includegraphics[width=4.1cm ]{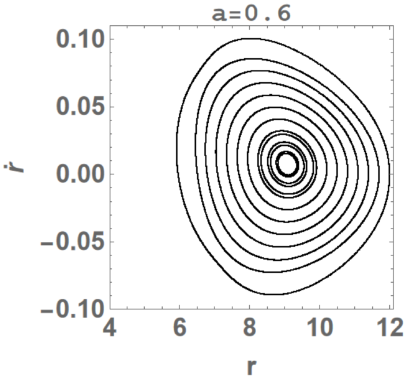}
	\caption{The change of the Poincar\'{e} section ($\theta = \frac{\pi}{2}$) with the swirling parameter $j$ (top, $B =10^{-4}$, $Q = a = 0.1$), magnetic field strength parameter $B$ (second, $j =10^{-6}$, $Q = a = 0.1$), electric charge parameter $Q$ (third, $j =10^{-5}$, $B =10^{-4}$, $a = 0.1$) and spin parameter $a$ (bottom, $j =10^{-5}$, $B =10^{-4}$, $Q = 0.1$) for the motion of particles around the dyonic Kerr-Newman black hole immersed in the Melvin-swirling universe, for which conical singularities have been removed. Here we set $M = 1$, $E = 0.95$ and $L = 2.4M$.}\label{fig12}
\end{figure}

Now we are in a position to investigate the chaotic motion of particles in the spacetimes free of conical singularities by using the Poincar\'{e} section. In Fig. \ref{fig12}, we plot the Poincar\'{e} sections with more motion orbits around the dyonic Kerr-Newman black hole in a Melvin-swirling universe, for which conical singularities have been removed.  We confirm the presence of chaotic behavior in the motion in this regular spacetime. It is shown that, as the swirling parameter $j$ and magnetic field strength parameter $B$ increase, the number of chaotic orbits and the chaotic region increase, but decrease as the electric charge parameter $Q$ and spin parameter $a$ increase. Obviously, the effects of parameters $j$, $B$, $Q$ and $a$ agree well with the findings in Figs. \ref{fig2}-\ref{fig4} and indicates that removing conical singularities will not influence the result.

\section{Chaotic motion of particles for the solution without both conical singularities and Dirac strings}

Just as shown in the solution (\ref{metric}), we can remove both the conical singularities and the Dirac strings by simultaneously considering the rescaling of the azimuthal coordinate $\phi \mapsto \frac{2\pi}{\delta \phi} \phi$
with the gauge transformation as $A_{\phi} \mapsto \frac{2\pi}{\delta \phi} A_{\phi} - \delta A_{\phi}$. Thus, an appropriate choice is \cite{DiPinto}
\begin{subequations}
	\label{reg-DKNMS}
	\begin{align}
		j & = \frac{ H \bigl(4 + 3 B^2 Z^2\bigr)}{4 Q Z^2} \,, \\
		a & = \frac{ Q B^3 Z^4}{2 M \bigl(4 + 3 B^2 Z^2\bigr)} \,, \label{reg-DKNMS-a}\\
		\delta \phi & = \frac{32 \pi Q^2 \bigl(4 + 3 B^2 Z^2\bigr)^2 \Bigl[ 16 + 8 B^2 \bigl( 3 H^2 + Q^2 \bigr) + B^4 Z^2 \bigl( 9 H^2 + Q^2 ) \Bigr]^{-1}}{Z^2  \Bigl(16 + 8 B^2 \bigl( 3 H^2 + 5 Q^2 \bigr) + B^4 Z^2 \bigl( 9 H^2 + 25 Q^2 \bigr) + 4 B^6 Q^2 Z^4 \Bigr)} \,, \\
		\delta A_{\phi} & = \frac{B Z^2 \Bigl[ 192 + 16 B^2 \bigl(23 H^2 + 19 Q^2 \bigr) + 12 B^4 Z^2 \bigl(19 H^2 + 15 Q^2 \bigr) + 45 B^6 Z^6 + 4 B^8 Q^2 Z^6 \Bigr]}{8 \bigl(4 + 3 B^2 Z^2 \bigr)^2} \,,
	\end{align}
\end{subequations}
which leads to the solution without both conical singularities and Dirac strings.

\begin{figure}[htbp!]
	\includegraphics[width=4.1cm ]{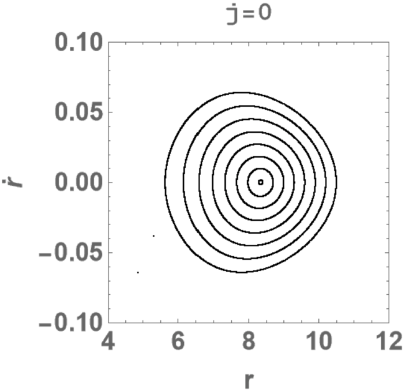}
	\includegraphics[width=4.1cm ]{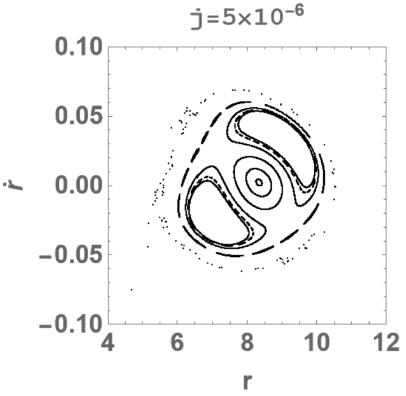}
	\includegraphics[width=4.1cm ]{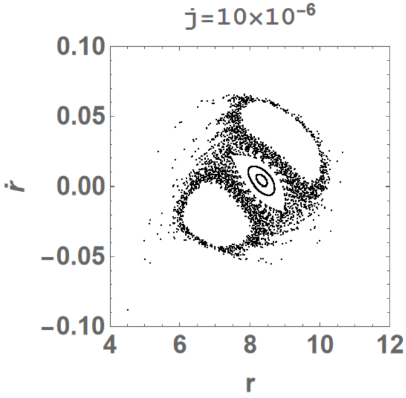}
	\includegraphics[width=4.1cm ]{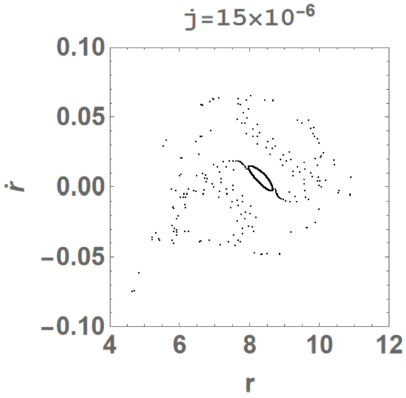}
	\includegraphics[width=4.1cm ]{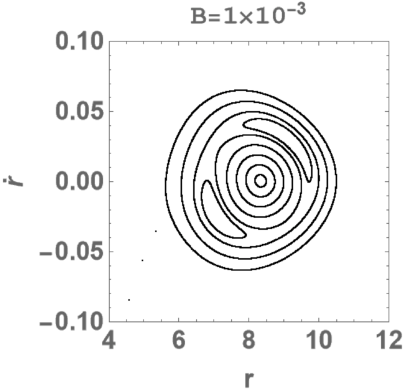}
	\includegraphics[width=4.1cm ]{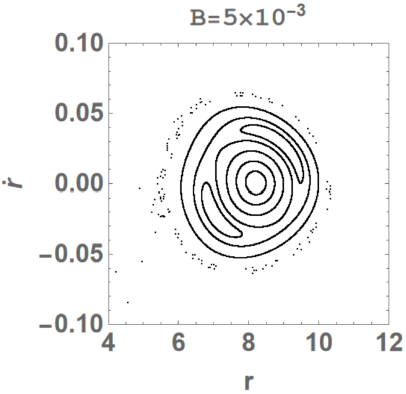}
	\includegraphics[width=4.1cm ]{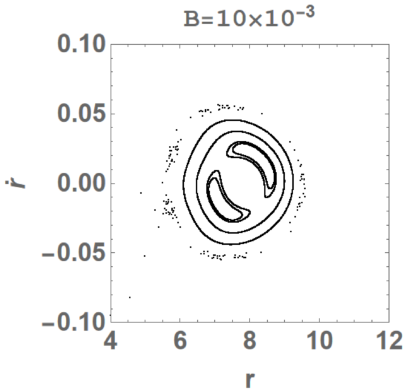}
	\includegraphics[width=4.1cm ]{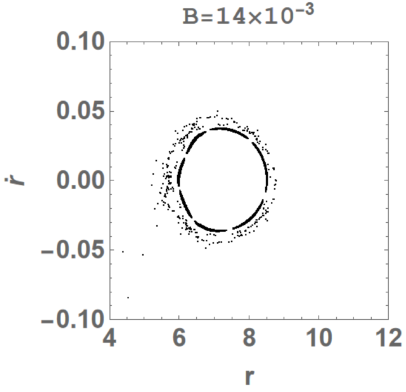}
	\includegraphics[width=4.1cm ]{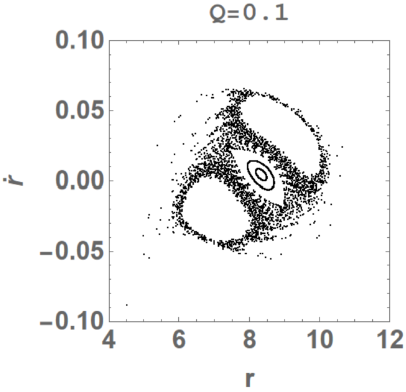}
	\includegraphics[width=4.1cm ]{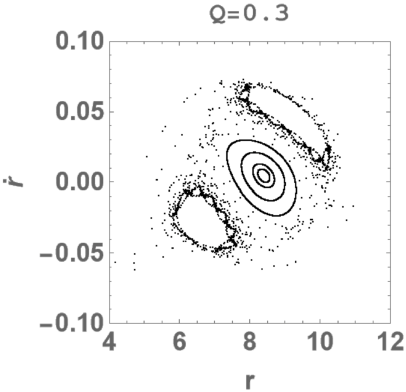}
	\includegraphics[width=4.1cm ]{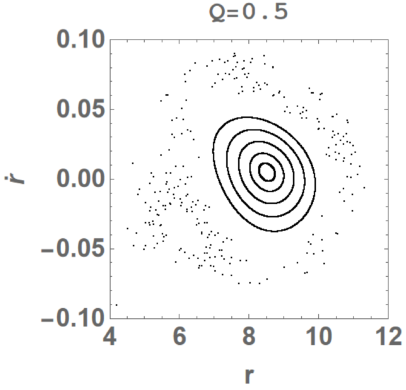}
	\includegraphics[width=4.1cm ]{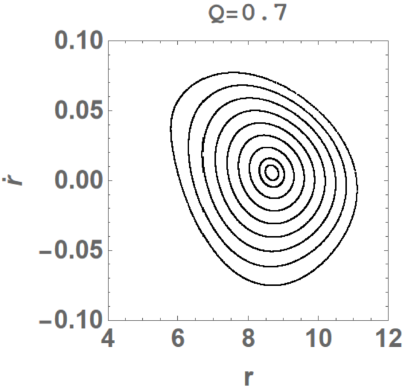}
	\caption{The change of the Poincar\'{e} section ($\theta = \frac{\pi}{2}$) with the swirling parameter $j$ (top, $B =10^{-4}$, $Q = 0.1$), magnetic field strength parameter $B$ (middle, $j = 10^{-6}$, $Q = 0.1$), and electric charge parameter $Q$ (bottom,  $j = 10^{-5}$, $B =10^{-4}$) for the motion of particles around the dyonic Kerr-Newman black hole immersed in the Melvin-swirling universe, for which both conical singularities and Dirac strings have been removed. Here we set $M = 1$, $E = 0.95$ and $L = 2.4M$.}\label{fig13}
\end{figure}

For the case where both conical singularities and Dirac strings have been removed, we present the change of the Poincaré section with varying values of the swirling parameter $j$, the magnetic field strength parameter $B$ and the electric charge parameter $Q$ for the dyonic
Kerr-Newman black hole immersed in the Melvin-swirling universe in Fig. \ref{fig13}. We observe that both the number of chaotic orbits and the chaotic region increase with the increase of $j$ and $B$, but decrease with the increase of $Q$, which is in good agreement with the result given in the Figs. \ref{fig2}, \ref{fig3}, \ref{fig4} and \ref{fig12}. Therefore, the effects of spacetime parameters are the same even both conical singularities and Dirac strings have been removed for the dyonic Kerr-Newman black hole immersed in the Melvin-swirling universe.

\end{document}